\documentclass[twocolumn,showpacs,preprintnumbers,amsmath,amssymb]{revtex4}
\usepackage{graphicx}
\usepackage{dcolumn}
\usepackage{bm}

\tighten

\newcommand{\de}{\delta}

\def\spose#1{\hbox to 0pt{#1\hss}}      
\def\ltapprox{\mathrel{\spose{\lower 3pt\hbox{$\mathchar"218$}}      
\raise 2.0pt\hbox{$\mathchar"13C$}}}      
\def\gtapprox{\mathrel{\spose{\lower 3pt\hbox{$\mathchar"218$}}      
\raise 2.0pt\hbox{$\mathchar"13E$}}}      
\def\inapprox{\mathrel{\spose{\lower 3pt\hbox{$\mathchar"218$}}      
\raise 2.0pt\hbox{$\mathchar"232$}}}

\newcommand{\be}{\begin{equation}}
\newcommand{\ee}{\end{equation}}
\newcommand{\bea}{\begin{eqnarray}}
\newcommand{\eea}{\end{eqnarray}}
\newcommand{\bean}{\begin{eqnarray*}}
\newcommand{\eean}{\end{eqnarray*}}
\newcommand{\bef}{\begin{figure}}
\newcommand{\eef}{\end{figure}}
\newcommand{\ba}{\begin{array}}
\newcommand{\ea}{\end{array}}

\newcommand{\bx}{{\bf x}}
\newcommand{\bk}{{\bf k}}
\newcommand{\bq}{{\bf q}}
\newcommand{\by}{{\bf y}}
\newcommand{\bbf}{{\bf f}}
\newcommand{\bF}{{\bf F}}
\newcommand{\bH}{{\bf H}}
\newcommand{\bu}{{\bf u}}

\newcommand{\bR}{{\bf R}}
\newcommand{\la}{\left<}
\newcommand{\ra}{\right>}
\def\lsim{\raise 0.4ex\hbox{$<$}\kern -0.8em\lower 0.62ex\hbox{$\sim$}}
\def\gsim{\raise 0.4ex\hbox{$>$}\kern -0.7em\lower 0.62ex\hbox{$\sim$}}

\begin{document}

\title{Force distribution in a randomly perturbed lattice of identical particles
with $1/r^2$ pair interaction}

\author{Andrea  \surname{Gabrielli}}    
\affiliation{Istituto dei Sistemi Complessi - CNR, Via dei Taurini 19, 
00185-Rome, Italy, \\
\& SMC-INFM, Physics Department, University 
``La Sapienza'' of Rome, 00185-Rome, Italy}
%
\author{Thierry \surname{Baertschiger}} 
\affiliation{Dipartimento di Fisica, Universit\`a ``La Sapienza'', 
00185-Rome, Italy,\\
\& Istituto dei Sistemi Complessi - CNR, Via dei Taurini 19, 
00185-Rome, Italy}
%
\author{Michael \surname{Joyce}}    
\affiliation{Laboratoire de Phyisique Nucl\'eaire et Hautes Energies, 
Universit\'e Pierre et Marie Curie - Paris 6, UMR 7585, 
Paris, F-75005, France} 
%
\author{Bruno \surname{Marcos}} 
\affiliation{ Dipartimento di Fisica, Universit\`a ``La Sapienza'', 
00185-Rome, Italy,\\
\& Istituto dei Sistemi Complessi - CNR, Via dei Taurini 19, 
00185-Rome, Italy}
  \author{Francesco \surname{Sylos Labini}} 
  \affiliation{``E. Fermi'' Center, Via Panisperna 89 A, Compendio del
Viminale, 00184 - Rome, Italy\\
\& Istituto dei Sistemi Complessi - CNR, Via dei Taurini 19, 
00185-Rome, Italy }

\begin{abstract}    
\begin{center}    
{\large\bf Abstract}    
\end{center}     
We study the statistics of the force felt by a particle
in the class of spatially correlated distribution of identical
point-like particles,
interacting via a $1/r^2$ pair force (i.e. gravitational or Coulomb),
and obtained by randomly perturbing an infinite perfect lattice.
In the first part we specify the conditions under which the
force on a particle is a well defined stochastic quantity.
We then study the small displacements approximation,
giving both the limitations of its validity, and, when it is 
valid, an expression for the force variance.
In the second part of the paper we extend to this class of particle 
distributions the method introduced by Chandrasekhar to study the
force probability density function in the homogeneous Poisson particle 
distribution. In this way we can derive an approximate expression
for the probability  distribution of the force over the full range of 
perturbations of the lattice, i.e., from very small (compared to the
lattice spacing) to very large where the Poisson limit is recovered.
We show in particular the qualitative change in the large-force 
tail of the force distribution between these two limits. 
Excellent accuracy of our analytic results is found on detailed
comparison with results from numerical simulations. These results
provide basic statistical information about the fluctuations 
of the interactions (i) of the masses in self-gravitating systems 
like those encountered in the context of cosmological N-body simulations,
and (ii) of the charges in the ordered phase of the One Component Plasma.
\end{abstract}    
\maketitle    
\date{today}    

\twocolumngrid  

\section{Introduction}

The study of the statistical properties of the force felt by a
particle in a gas, and exerted by all the other particles of
the system through pair interactions, can provide useful insights
into the thermodynamics or dynamics of physical systems in many 
different contexts and applications.  Some important examples are given by: (i)
the distribution of the gravitational force in a spatial distribution
of point-like masses in cosmological and astrophysical applications
\cite{chandra,star,fractal}, (ii) the statistics of molecular and
dipolar interactions \cite{dipole} in a gas of particles, (iii) the
theory of interacting defects in condensed matter physics
\cite{dislo}, and (iv) the contact force distribution in granular
materials \cite{granular1,granular2}.

The seminal work in this field is due to Chandrasekhar \cite{chandra}
and deals mainly with the probability density function (PDF) of the
gravitational force in a homogeneous spatial Poisson distribution of
identical point-like masses. Specifically the PDF of the force is
found to be given exactly by the Holtzmark distribution, which is a
three-dimensional (3D) fat tailed L\'evy distribution \cite{levy}.
In \cite{star,gauss-poisson,dipole,libro} approximated
generalizations, in different physical contexts, of this approach can
be found for more complex point-like particle systems obtained by
perturbing a homogeneous Poisson particle distribution.

Here we present a study of the probability distribution of the total
gravitational (or Coulomb) force for a specific class of spatial particle
distributions (i.e. point processes): three-dimensional {\em
shuffled lattices}, i.e., lattices in which each particle is randomly
displaced from its lattice position with a PDF of the displacement
$p(\bu)$.  This study can be very useful for applications in both
solid state physics (e.g. in the case of Coulomb or dipolar pair
interaction) \cite{dipole} and in astrophysics and cosmology.
In the latter context, specifically, large ``N-body'' numerical
simulations of self-gravitating particles are an essential instrument
in the study of the problem of structure formation in the universe
starting from small initial mass fluctuations. These simulations
are performed usually starting from initial conditions which
are suitably perturbed simple lattices \cite{n-body,ic}. We will
discuss briefly in our conclusions the kinds of questions 
which may be addressed in this context using the formal results and
methods developed in this paper.

Quite generally Chandrasekhar \cite{chandra} showed that the 
gravitational force acting on each particle
can be seen as due to the superposition of two different
contributions: the former (a sort of {\em fluid} term) is due to the
system as a whole, and the latter is due to the influence of the
immediate neighborhood of the particle and therefore depends
on how a spatial mass distribution (e.g. a fluid) is represented through
a point-like particle system.  
The former is a smoothly varying function of
position while the latter is subject to relatively rapid
fluctuations. We will show that this is true also for the case of a
shuffled lattice, even though some important difference with respect
to the case of the homogeneous Poisson case are present when the
shuffling is very small due the extreme uniformity of the particle
distribution even at small scales. When, instead, the shuffling is
sufficiently large we recover approximately the same behavior as in 
the Poisson case.

The rest of the paper is organized as follows:

\begin{itemize}

\item Sect.~\ref{sec2}: The principal one and two point correlation properties 
of point processes and in particular of a shuffled lattice are briefly 
reviewed;
\item Sect.~\ref{sec3}: We give the statistical definition of the
global gravitational force acting on each particle specifying the
conditions under which this is a well defined stochastic quantity. 
In particular
we discuss the problem posed by the infinite volume limit and the necessity
of introducing a compensating uniform negative background mass density
for the statistical definiteness of the gravitational force in this
limit.  Here we point out also that this study and the definitions
given are valid in both the cases of gravitational and 
Coulomb pair interaction;
\item Sect.~\ref{sec4}: We study the stochastic total gravitational
force acting on a particle to linear order in the random
displacements. In this way we identify two different contributions to
this force: the former comes from the displacements of the point-like
sources keeping the particle on which we calculate the force at its
original lattice position, and the latter comes from the displacement
of this particle with respect to the rest of the lattice.  The first
contribution is dominated by the particles in the neighborhood of the
particle on which we calculate the force, while the second can be seen 
as a force due to the system as a continuous mass distribution; 
\item Sect.~\ref{sec5}: Here the variance of the gravitational
force is calculated in the above approximation and its meaning is
discussed in relation to the form $p(\bu)$ of the displacement PDF.
In particular we find the fundamental difference of the behavior
between the two cases in which the particle displacements are limited
or not to the elementary lattice cell;
\item Sect.~\ref{sec6}: In this section, in order to go beyond a study
of the statistical properties of the gravitational force limited to
the first and second moment, we briefly report some previously known 
results about the PDF of the force in two different situations: (i) the first
is the exact solution of Chandrasekhar for the case of a three-dimensional 
homogeneous Poisson particle distribution, and (ii) the second concerns
the PDF of the total force in one-dimensional (1D) shuffled lattices of
particles interacting via generic power law pair-interactions;
\item Sect.~\ref{sec7}: We use the results exposed in the previous
section to develop and discuss an approximate evaluation {\em \`a la}
Chandrasekhar of the PDF (i.e. of all the moments) of the total force
acting on a given particle. This provides much useful information
about the exact PDF of the force.  The approximation becomes better
and better when the typical permitted displacements of the particles
approach and go beyond the limit of the elementary lattice cell;
\item Sect.~\ref{sec8}: In this section we perform the comparison of
the theoretical and analytical results found in the preceding
sections with the results obtained by numerical simulations.
The agreement is shown to be very good in general despite the fact that
the three-dimensional problem of the gravitational force in a shuffled
lattice is not exactly solvable;
\item Sect.~\ref{sec9}: Finally we summarize the main results of
this work and draw some concluding remarks on their utility.

\end{itemize}

\section{Statistical properties of a shuffled-lattice}
\label{sec2}

Let us introduce some basic notation that will be useful in the rest
of the paper. Given a spatial distribution of particles in a cubic volume 
$V$ (we indicate with $V$ both the space region in which the system is defined
and its volume; in this paper we are interested in the limit 
$V\rightarrow \infty$) with equal
mass $m$ (i.e. a so called {\em point process} \cite{libro,intro-to-pp}), the
microscopic mass density function is:
\be
\rho(\bx)=m\sum_i \delta(\bx-\bx_i)\,,
\label{2-1}
\ee 
where $\delta(\bx)$ is the 3D Dirac delta function,
$\bx_i$ is the position of the $i^{th}$ particle of the system and the
sum is over all the particles of the system.  
Clearly the microscopic number density
is simply given by $n(\bx)=\rho(\bx)/m$.
Let us suppose that $n_0>0$ is the average number
density. Consequently, the average mass density is simply
$\rho_0=n_0m$.  The power spectrum (PS) of such a system is defined as 
\be
{\cal P}(\bk)\equiv \lim_{V\rightarrow \infty} \frac{\la|\delta \hat n
(\bk;V)|^2\ra}{n_0^2V}\equiv \lim_{V\rightarrow \infty}
\frac{\left<|\delta \hat \rho (\bk;V)|^2\right>}{\rho_0^2V}\,,
\label{2-2}
\ee where $\la...\ra$ indicates the average over all the realizations
of the point process, and
\[
\delta \hat n (\bk;V)=\int_V d^3x\,e^{-i\bk\cdot\bx}[n(\bx)-n_0]
\] 
is the Fourier integral in the volume $V$ of the number density
contrast $[n(\bx)-n_0]$
\footnote{In the limit $V\rightarrow \infty$, we adopt in general the
following usual normalizations for the Fourier transform: ${\cal
F}[g(\bx)]= \int d^3x\,e^{-i\bk\cdot\bx} g(\bx)$ is the Fourier
transform of the function $g(\bx)$, and ${\cal F}^{-1}[\hat
g(\bk)]=\frac{1}{(2\pi)^3}\int d^3k\,e^{i\bk\cdot\bx}\hat g(\bk)$ the
inverse Fourier transform.}.  Note that ${\cal P}(\bk)$ does not
depend on $m$.  Therefore without loss of generality we fix $m=1$ for
sake of simplicity and use $n(\bx)$ for both the number and the mass
density function.

In general an initial particle distribution can be perturbed by
applying a stochastic displacement to each  particle of the
system (see Fig.~\ref{config}).
In particular a perturbed lattice is built by applying such a
displacement to
each particle initially belonging to a regular lattice (e.g. simple
cubic). If $\bR$ is the generic lattice site 
the density function $n(\bx)$ for a regular lattice reads
\be
n(\bx)=\sum_{\bR} \delta(\bx-\bR)\,,
\label{2-3}
\ee
where the sum is over all the lattice sites.
If, moreover, $\bu(\bR)$ is the displacement applied to the particle 
initially at $\bR$, the function $n(\bx)$ for such a SL will be
\be
n(\bx)=\sum_{\bR} \delta[\bx-\bR-\bu(\bR)]\,.
\label{2-4}
\ee 
In Fig.~\ref{sl-3d} we present a typical configuration of a 3D
perturbed cubic lattice projected on one of the lattice planes.
The perturbed lattice is said to be {\em uncorrelated} if the
displacements applied to different particles are taken
independent of each other. We call such a system a {\em shuffled}
lattice (SL). This means that
$p^{(2)}[\bu(\bR),\bu(\bR')]=p[\bu(\bR)]p[\bu(\bR')]$, where
$p^{(2)}[\bu(\bR),\bu(\bR')]$ is the {\em joint} PDF of the 
displacements applied to two different particles
respectively initially at $\bR$ and $\bR'$, and $p(\bu )$ the PDF of a
single displacement.  
\bef
\includegraphics[height=6.5cm,width=8cm,angle=0]{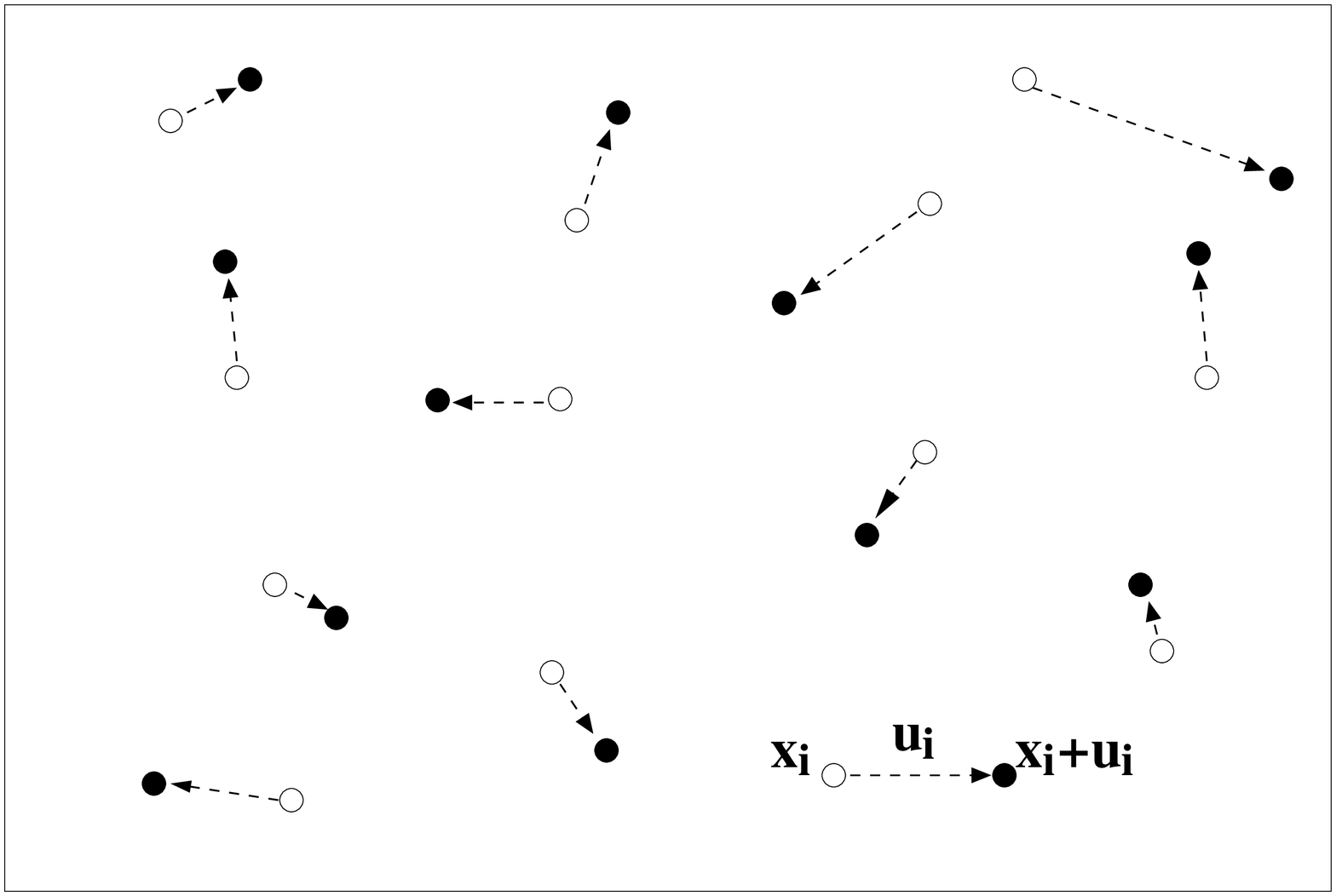}
\caption{The figure presents a pictorial view of the effect of a
stochastic displacement field on a spatial particle distribution in
$2D$.  The particles move, through the displacements (dashed arrows),
from the old positions (empty circles) to the new ones (black circles).
\newline
\label{config}}
\eef
Without loss of generality in the following we limit the discussion to
the symmetric case in which $p(\bu)=p(-\bu)$.
Clearly the average over all the possible
realizations of the displacement field coincides with the average over
all the possible realizations of the point process. Therefore, we will
use the notation
\[\la g(\bu_1,...,\bu_l)\ra=\int d^3u_1...d^3u_l\,g(\bu_1,...,\bu_l)\,
p(\bu_1)...p(\bu_l)\]
for the average of any function of the $l$ displacements applied 
respectively to $l$ different points.

\bef
\includegraphics[height=8cm,width=8cm,angle=0]{fig2.eps}
\vspace{0.6cm}
\caption{The figure provides the projection on the $x-y$ plane of a 3D
{\em shuffled} simple cubic lattice of $16^3$ particles in a cubic
unitary volume. In this case $p(\bu)=\prod_{i=1}^3 f(u_i)$ where
$f(u_i)= \theta(\Delta-|u_i|)/(2\Delta)$ (as in the simulations
considered in Sect.~\ref{sec8}) with $2\Delta=\ell/4$ (i.e. each
particle is displaced well inside its elementary lattice cell).
\label{sl-3d}}
\eef

By calling $\hat p(\bk)$ the {\em characteristic function} of a
single stochastic displacement, i.e., its Fourier
transform (FT), we can write the PS
of the final point process as \cite{displa}
\be
\label{ps_sl}
{\cal P}(\bk)=
\frac{1}{n_0}\left(1-|\hat p(\bk)|^2\right)+(2\pi)^3|\hat p(\bk)|^2
\sum_{{\bf H} \ne {\bf 0}}\delta(\bk-{\bf H})\,,
\ee
where ${\bf H}$ is the generic vector of
the reciprocal lattice \cite{ashcroft}
of the real space lattice giving the initial particle configuration, and 
\[
{\cal P}_{\mbox{in}}(\bk) = (2\pi)^3\sum_{{\bf H} \ne {\bf
0}}\delta(\bk-{\bf H})
\] 
is the PS of this initial regular configuration.  A $1d$ example of
such a PS is given in Fig.~\ref{sl-k2} in which the exact formula
(\ref{ps_sl}) is compared with the PS evaluated in numerical
simulations of the SL, showing a perfect agreement.  Note that if the
applied displacement field is isotropic, $p(\bu)$ depends only on
$u=|\bu|$ and $\hat p(\bk)$ only on $k=|\bk|$.  
\bef [tbp]
\includegraphics[height=7cm,width=8.5cm,angle=0]{fig3.eps}
\vspace{0.2cm}
\caption{Comparison between the PS $S(k)=n_0^2{\cal P}(k)$ measured in
numerical simulations (circles) of a 1D SL with the theoretical
prediction (continuous curve) given by Eq.~(\ref{ps_sl}) in the case
in which the one displacement PDF $p(u)=\theta(\Delta-|u|)/(2\Delta)$
with $\delta=\Delta/\ell=1/50$.  In order to represent appropriately
the modulated delta functions contribution to the PS $S(k)$, we have
normalized their amplitude to a value $10^2$ for the unperturbed
lattice.  The numerical result is obtained by averaging over $10^3$
realizations of the same SL and the agreement with the
theoretical prediction is excellent.
\label{sl-k2}}
\eef

The connected
two-point correlation function \cite{libro,GJS}, 
also called covariance function, is defined in
general as 
\[\tilde\xi(\bx,\bx')=\frac{\left<n(\bx)n(\bx')\right>}{n_0^2}-1\,,\]
In the case of statistical translational invariance it 
is a function only of the vector distance $\bx-\bx'$, i.e.,
$\tilde\xi(\bx,\bx')=\tilde\xi(\bx-\bx')$ and satisfies the relation
\[
\tilde
\xi(\bx)={\cal F}^{-1}[{\cal P}(\bk)] \;.
\] 
However, as the initial particle configuration is a regular lattice, 
for a SL there is not full translational invariance. In this case ${\cal
F}^{-1}[{\cal P}(\bk)]$ can be seen as the average of
$\tilde\xi(\bx_0+\bx,\bx_0)$ over $\bx_0$ in an elementary lattice cell
\footnote{Note that this means that when there is not
statistical translational invariance ${\cal P}(\bk)$ contains less
information than $\tilde\xi(\bx,\bx')$.}.  Note that for any particle 
distribution
\be
\tilde\xi(\bx)={\cal F}^{-1}[{\cal P}(\bk)]=\frac{\delta(\bx)}{n_0}+
\xi(\bx)\,,
\label{off} 
\ee
where $\delta(\bx)/n_0$ is the singular ``diagonal part'' 
of the covariance function due to the fact that the microscopic density has
an infinite discontinuity on any particle, and $\xi(\bx)$ is the
smooth ``off-diagonal'' part giving the real two-point correlation
between different particles. One important property is $\xi(\bx)\ge -1$
for all $\bx$. This comes from the fact that the average density
of particles seen by any particle of the system cannot be negative.
The off-diagonal part $\xi(\bx)$ vanishes 
identically only for the homogeneous Poisson point process
\cite{libro,intro-to-pp}. 

It is evident from Eq.~(\ref{ps_sl}) that the random shuffling
$\{\bu(\bR)\}$ in general does not erase completely in the PS 
${\cal P}(\bk)$ the
presence of the so-called {\em Bragg peaks} (i.e. the sum of delta
functions) for each reciprocal lattice vector $\bH\ne 0$,
of ${\cal P}_{\mbox{in}}(\bk)$, but only modulates their amplitude
proportionally to $|\hat p(\bH)|^2$, and adds a continuous
contribution typical of stochastic particle distributions $(1-|\hat
p(\bk)|^2)/n_0$. Around $k=0$ (more precisely in the whole {\em first
Brillouin zone} \cite{ashcroft})
${\cal P}_{\mbox{in}}(\bk)=0$ identically (i.e. we can say that around $k=0$
${\cal P}_{\mbox{in}}(\bk)\sim k^\infty$). Thus, from this point of view,
regular lattices can be seen as the
class of the most {\em superhomogeneous} particle distributions
\cite{libro,GJS}, i.e., of the particle distributions with a PS satisfying
${\cal P}(\bk)\sim k^a$  with $a>0$ at small $k$
which is equivalent to the condition $\int d^3x\, \tilde\xi(\bx)=0$.  
As is clear from Eq.~(\ref{ps_sl}), in this region
${\cal P}(\bk)$ is determined by only the behavior of the displacement
characteristic function $\hat p(\bk)$. In particular, even though the
lattice is strictly anisotropic, this implies that if the displacement
field is statistically isotropic the final SL
has isotropic mass fluctuations at large scales (i.e. for
$k\rightarrow 0$). By assuming $p(\bu)=p(u)$, it is now
simple to show \cite{displa} that for any PDF at small $k$ we have
\be
\hat p(k)=1-Ak^\alpha +o(k^\alpha)\,,
\label{2-5}
\ee
where $\alpha=2$ and $A=\overline{u^2}/6$ if $\overline{u^2}$ is finite
(where $\overline{f(\bu)}=\int d^3u\,p(\bu)f(\bu)$), 
while $\alpha=\beta-3$ and $A>0$ if $p(u)$ decays as $u^{-\beta}$ at large $u$
with $\beta<5$ (note that $\beta>3$ for any PDF $p(u)$ in three dimensions),
and therefore $\overline{u^2}$ is infinite.

In this paper we focus our attention on the class of SL, where, as 
written above, the applied displacements are statistically uncorrelated.

\section{Definition of the gravitational force on a particle
in a perturbed lattice}
\label{sec3} 

Let us now consider basic questions concerning the 
definition 
of the gravitational force acting on a particle in an infinite perturbed
lattice.  As
aforementioned we assume that: (i) all particles have mass $m=1$, (ii)
the average number density is $n_0=\ell^{-3}$, where $\ell$ is the lattice
spacing, (iii) the microscopic number density is given by
Eq.~(\ref{2-4}), and (iv) we choose the units so that the gravitational
constant $G=1$.
Let us suppose for simplicity that the initial position of
the particle on which we evaluate the gravitational field is the
origin of coordinates (i.e. $\bR=0$).  The total gravitational
force acting on it is:
\be 
\bF=\sum_{\bR\ne
0}\frac{\bR+\bu(\bR)-\bu(0)}{|\bR+\bu(\bR)- \bu(0)|^3}\,,
\label{3-1}
\ee 
where the sum is over all the particles initially at the lattice
sites $\bR\ne 0$.
The same sum can be simply rewritten as
\be
\bF=
\int_V d^3x\, n'(\bx)\frac{\bx-\bu(0)}{|\bx-\bu(0)|^3}\,,
\label{3-1-b}
\ee
where $V$ is the system volume, and 
\be
n'(\bx)=n(\bx)-\delta[\bx-\bu(0)]\,,
\label{n-primo}
\ee
with $n(\bx)$ given by Eq.~(\ref{2-4}).

Note that by simply inverting the sign of the pair
interaction, and therefore of the total force from attractive to
repulsive, and substituting the identical masses with identical
charges, Eq.~(\ref{3-1}) gives the total repulsive Coulomb
force $\bF_C$ in a perturbed Coulomb lattice \cite{pines} on the
point-charge at $\bu(0)$ and generated by the all other point-charges
at $\bR+\bu(\bR)$:
\be
\bF_C=\sum_{\bR\ne
0}\frac{\bu(0)-\bR-\bu(\bR)}{|\bR+\bu(\bR)- \bu(0)|^3}=
\int_V d^3x\, n'(\bx)\frac{\bu(0)-\bx}{|\bx-\bu(0)|^3}
\,.
\label{3-1-c}
\ee  
Consequently, all the results in this
paper can be directly applied to the statistics of the repulsive
force in a shuffled Coulomb lattice.

We are interested in the limit $V\rightarrow \infty$ of
Eq.~(\ref{3-1-b}), where we mean by this limit that the volume
$V$ tends to the whole of $\mathbb{R}^3$.  It is well
known in different physical contexts \cite{ashcroft,binney} that,
if $n_0>0$, the infinite volume limit of lattice sums such as those
in Eq.~(\ref{3-1}) or (\ref{3-1-c}) is in fact not well defined because 
these sums are only conditionally convergent in the limit 
$V\rightarrow\infty$, i.e.,
their result depends on the order in which the single terms are summed.
In many physical applications, however, as in the case of the
Coulomb lattice \cite{pines}, this sum is regularized automatically by
the presence in the physical system of a uniform background charge
density with opposite sign with respect to that of the point-charges
and such that there is overall charge neutrality.  Once the
attractive force $\bF_b$ of the background on the point-charges is
taken into account, by adding the corresponding term
\[\bF_b(\bu_0)=n_0\int_V d^3x\,\frac{\bx-\bu_0}{|\bx-\bu_0|^3}\] 
to Eq.~(\ref{3-1-c}),  then the total Coulomb force
acting on a given point-charge is finite and its value is independent
of the way in which the infinite volume limit is taken \footnote{
This is true provided $V$ is a compact set of $I\!\!R^3$, containing
both the point-charges and the uniform opposite charged background.}.
To clarify this point, let us consider the following system: a
density of point-charges $n(\bx)$ given by Eq.~(\ref{2-1}) with $m=1$ embedded
in a uniform background charge density $-n_0=-\left<n(\bx)\right>$. 
Therefore the local charge density at the
point $\bx$ will be $\delta n(\bx)=n(\bx)-n_0$.  The
force acting on a probe charge at the point $\by$ of the space and generated 
by the total charge in the volume $V$ around it will be:
\be
\bF(\by;V)=\int_V d^3x\,\delta n(\bx)\frac{\by-\bx}{|\by-\bx|^3}\,.
\label{3-1-d}
\ee
 Let us now assume that the PS of $n(\bx)$, and therefore of $\delta n(\bx)$,
has the behavior ${\cal P}(\bk)\sim k^a$ at small $k$,
with $a>-3$ (in three dimensions). Under these hypotheses 
(see e.g. \cite{displa}) it is simple to show that 
\[\left<\left|\int_V d^3x\,\delta n(\bx)\right|^2\right>\sim V^{b/3}\]
with $b=3-a$ if $a<1$ and $b=2$ if $a\ge 1$. From Eq.~(\ref{3-1-d}) we 
then expect in general that $\bF(\by)=\bF(\by;V\rightarrow+\infty)$ is 
a well defined stochastic quantity
(i.e. a stochastic quantity whose statistics is well defined and not
depending on how the compact volume $V$ is sent to infinity) if
$a>-1$ \footnote{Note,
however, that the {\em difference} of the force between any two points
of the space is a well defined stochastic quantity for any
statistically homogeneous particle distribution with $a>-3$, i.e.,
when the mass density PS is well defined for $k\rightarrow 0$
(i.e. integrable). In systems with $-3<a\le -1$ the force acting on
the center of mass of any subregion
would be divergent in the infinite volume limit,
but this is not relevant for the evolution of the relative distance of
the particles.}. Indeed for these values of $a$ fluctuations in the density
contrast $\delta n(\bx)$ generate in the infinite volume limit
quadratic fluctuations in the force $\bF$ which are size independent.
This is, in particular, the case for the SL we consider in this paper,
for which we have shown that $a>0$ for any displacement PDF $p(\bu)$
(see Eq.~(\ref{2-5})). Given that the limit $V\rightarrow\infty$ of
Eq.~(\ref{3-1-d}) does not depend on the way in which the limit is
performed, we can choose for simplicity to take the volume $V$
symmetric with respect to the point $\by$ where the force is computed.
In such a volume the contribution to the force $\bF(\by;V)$ 
from the background vanishes by symmetry.
Consequently, with this choice of the volume $V$,
the force $\bF(\by)$ coincides with the limit of the sum (\ref{3-1-d}) 
with $\de n(\bx)$ replaced by $n(\bx)$ (i.e. summing only
over the particles).

This in particular implies that for the SL particle distribution
we will study here, in the presence of an oppositely charged 
neutralising background, we can evaluate the well defined global 
force $\bF$ acting on the point-charge in $\bu(0)$
simply using Eq.~(\ref{3-1-c}) where the sum is over all the
charges contained in a sphere $S[r;\bu(0)]$ of radius $r$ centered on
$\bu(0)$ and then taking the limit $r\rightarrow \infty$.  Note that,
on the other hand, the limit $r\rightarrow\infty$ of Eq.~(\ref{3-1-c})
using instead the sphere $S[r;0]$ centered on the point $0$, where the
considered particle was {\em before} the displacement, does not give
the full force $\bF$ on the point $\bu(0)$. To obtain it the
background contribution, which now is different from zero, must be
added.

In the rest of this paper, for convenience, we will choose to take the 
infinite volume limit in the former way, i.e., symmetrically in
spheres about the considered particle, also in the treatment 
of the gravitational problem, i.e., for Eq.~(\ref{3-1}).  
It is in fact is possible to show that the presence of an analogous 
background with a negative mass density
(now exerting a repulsive force on the massive particles) comes
out naturally when the motion of a particle is described in comoving
coordinates starting from the exact equations of general relativity in
a quasi-uniform expanding Universe (see e.g. \cite{Peebles80}).  In
pure Newtonian gravity, instead, such a background does not exist and
has to be introduced artificially to regularize the problem ({\em
Jean's swindle}, see \cite{binney}).  The results given here
for the statistics of the force $\bF$ have thus to be understood 
as strictly valid in the presence of such a compensating background.

In \cite{bsl04} some of us (TB and FSL) have given a simple estimation 
of the contribution of the first six nearest
neighbors (NN) of the particle to Eq.~(\ref{3-1}) for a simple cubic lattice.  
We show now that
instead the sum of Eq.~(\ref{3-1}) on a sphere of radius $r$ around the
central particle can be seen as the sum of two different
contributions: the first ``asymmetric'' one is due to the
self-shuffling $\bu(0)$ of the center particle from the initial $\bR=0$, 
and can be seen as induced by the system as a whole. The second ``symmetric''
term is due to the shuffling of all the other particles, and 
is dominated by the contribution of the first six NN.

\section{The small displacements behavior of the force} 
\label{sec4}

In this section we will give the approximate expression of $\bF$
obtained through a linearization in the particle displacements. The
statistical meaning and the limitations on taking averages of powers of
this linearized expression will be discussed in the next section.
To simplify our computation, but without loss of generality of
the results, we limit the discussion to those lattices with a cubic symmetry:
i.e., simple cubic, body centered cubic, and face centered cubic lattices.
 
As shown above, let us rewrite Eq.~(\ref{3-1}) as
\be
\label{m1}
\bF = \lim_{r \rightarrow \infty} \int_{S[r;\bu(0)]} n'(\bx)
\frac{\bx-\bu(0)}{|\bx-\bu(0)|^3}d^3x 
\ee 
where the integral is over the sphere $S[r;\bu(0)]$ defined
above, and $n'(\bx)$ is given by Eq.~(\ref{n-primo}).
In this section we are interested in the linear contribution
in the displacement field $\bF^{(l)}$ to Eq.~(\ref{m1})
for small displacements.
Note that 
\[d\bF(\bx)=\, n'(\bx) 
\frac{\bx-\bu(0)}{|\bx-\bu(0)|^3}d^3x\]
is the force contribution coming from the volume element $d^3x$ 
around $\bx$. Therefore we can rewrite Eq.~(\ref{m1}) as
\be
\label{m1b}
\bF = \lim_{r \rightarrow \infty} \bF[r;\bu(0)]=\lim_{r \rightarrow \infty} 
\int_{S[r;\bu(0)]} d\bF(\bx)
\ee
We now write
\be
\label{m2}
\int_{S[r;\bu(0)]} d\bF(\bx) = \int_{S(r;0)} d\bF(\bx) +
\int_{\delta S[r;\bu(0)]} d\bF(\bx) \ee 
where $S(r;0)$ is the sphere of radius $r$ centered on the lattice
point $\bR=0$, and the integration over 
$\delta S[r;\bu(0)]$ means the integration over the portion of
$S[r;\bu(0)]$ not included in $S(r;0)$ minus that one over the portion of
$S(r;0)$ not included in $S[r;\bu(0)]$ \footnote{This coincides with
the integration over the sphere $S[r;\bu(0)]$ minus the integration on 
the sphere $S(r;0)$.}.
Note that these portions of sphere have both the same volume, 
which is proportional to $|\bu(0)|$.
Consequently, we expect that this second integral will give a 
force contribution of order $\bu(0)$.

Let us start by evaluating the first term in Eq.~(\ref{m2}):
\[\bF_S(r)\equiv \int_{S(r;0)} d\bF(\bx) \]
to first order in the displacements.  
It can be
written as Eq.~(\ref{3-1}) where the sum is over all the
particles contained in the sphere $S(r;0)$ with the exception of the
one at $\bu(0)$. We thus denote this kind of sum by
$\sum_{\bR\ne 0}^{R\le r}$, i.e.,
\be
\bF_S(r)=\sum_{\bR\ne 0}^{R\le r}
\frac{\bR+\bu(\bR)-\bu(0)}{|\bR+\bu(\bR)-\bu(0)|^3}\,.
\label{3-2-0}
\ee
We are interested in the contribution $\bF_S^{(l)}(r)$
to this quantity at linear order in the displacement field.
Performing a Taylor expansion of Eq.~(\ref{3-2-0}) to first 
order in the (relative) displacements, we can write:
\be
\bF_S^{(l)}(r)=\sum_{\bR\ne 0}^{R\le r}\bbf_{\bR}\,,
\label{3-2}
\ee
with
\begin{equation}
\label{force_expansion}
\bbf_{\bR}= \frac{3[\bu(0)\cdot\hat{\bR}]\hat{\bR}-\bu(0)}{R^3}
+\frac{\bu(\bR)-3[\bu(\bR)\cdot\hat{\bR}]\hat{\bR}}{R^3}
\end{equation}
where $\hat\bR=\bR/R$ is the unit vector in the direction of the
lattice site $\bR$.  
The quantity $\bF_S^{(l)}(r)$ will be a good approximation to 
$\bF_S(r)$ if all the (relative) displacements are sufficiently 
small (see next section).
Eqs.~(\ref{3-2}) and (\ref{force_expansion}) show that, 
at first order, the contribution to the
force separates into a part due to the displacement of the
particle on which we calculate the force, and a part due to the
displacement of the particle originally at $\bR$. Let us write
\begin{equation}
\bbf_\bR= \bbf_\bR^o + \bbf_\bR^s 
\end{equation}
with 
\begin{eqnarray}
&&\bbf_\bR^o=\frac{-\bu(0)+3 [\bu(0)\cdot\hat{\bR}]}{R^3}
 \\
&&\bbf_\bR^s=\frac{\bu(\bR)-3[\bu(\bR)\cdot\hat{\bR}]\hat{\bR}}{R^3}
\;.
\end{eqnarray}
It is simple to show, for the class of lattices we consider,
that 
\begin{equation}
\label{3-3}
\sum_{\bR\ne 0}^{R\le r} \bbf_\bR^o\equiv 0 
\end{equation}
when the sum is taken up to include all the lattice sites $\bR\ne 0$
distributed symmetrically with respect to $\bR=0$ up to a given
distance.  The key points to show this are: (i) to rewrite
the sum in Eq.~(\ref{3-3}) as a sum over all the permitted values of
$R=|\bR|\le r$ of the sums over all the sites with fixed $R$;  (ii)
to consider that for the sub-sum over all the sites $\bR$, with
fixed $R=|\bR|$ we can write:
\[
\sum_{\bR}^{|\bR|=R} \bbf^o_\bR= 
\frac{1}{R^3}
\sum_{\bR}^{|\bR|=R} \left(-\bu(0)+3 
[\bu(0) \cdot\hat{\bR}]  \hat{\bR} \right)\,;
\]
(iii) and to note that 
\[3\sum_{\bR}^{|\bR|=R} [\bu(0)\cdot\hat{\bR}]  \hat{\bR} 
=\sum_{\bR}^{|\bR|=R} \left(\hat{\bR}\cdot\hat{\bR}\right) \bu(0) =
\sum_{\bR}^{|\bR|=R} \bu(0)\]
from which Eq.~(\ref{3-3}) follows.
Therefore we can conclude that 
\be
\bF_S^{(l)}(r)=\sum_{\bR\ne 0}^{R\le r}\frac{\bu(\bR)-3[\bu(\bR)\cdot 
\hat\bR]\hat\bR}{R^3}\,.
\label{3-3-b}
\ee
and we will call simply
\[\bF^{(l)}_S=\lim_{r\rightarrow\infty}\bF_S^{(l)}(r)\,.\]

Let us consider now the second term in Eq.~(\ref{m2}) which we call
$\bF_{A}(r)$:
\be
\label{m5}
\bF_{A}(r) = \int_{\delta S[r;\bu(0)]} d^3x
\frac{n'(\bx) (\bx-\bu(0))} {|\bx-\bu(0)|^3}\,. 
\ee
Since we wish to evaluate the
above integral to linear order in $\bu(0)$, and the volume of
integration is already of order $|\bu(0)|$, we can substitute $n'(\bx)$ with
its average (the point $\bu(0)$ is not included in the volume of
integration) $n_0=\ell^{-3}$ and put $\bu(0)=0$ in the integrand.  It is then
straightforward to show that, taking the limit
$r\rightarrow\infty$ and working to first order in $\bu(0)$, we have 
\be 
\bF_A
\simeq \bF^{(l)}_A=\frac{4\pi}{3} n_0 \bu(0)\,,
\label{m6}
\ee
where we have simply called $\bF_A=\lim_{r\rightarrow\infty}\bF_A(r)$
and $\bF^{(l)}_A$ its approximation at linear order in the displacement.
Note that, as discussed in the previous section, this quantity can be
seen as the force exerted by the negative background contained in the
sphere $S(r;0)$ on the particle at $\bu(0)$.

Let us summarize before proceeding further in the next section.
We have now seen that, to first order in the displacements, we
can write 
\begin{widetext}
\be
\label{3-4}
\bF^{(l)}=\bF_A^{(l)}+\bF_S^{(l)}
=\frac{4\pi}{3}n_0\bu(0)+\lim_{r\rightarrow
\infty} \sum_{\bR\ne 0}^{R\le r} \frac{\bu(\bR)-3[\bu(\bR)\cdot 
\hat\bR]\hat\bR}{R^3}\,,
\ee
\end{widetext}
where the sum is taken in the sphere $S(r;0)$. As explained in
Sect.~\ref{sec3}, Eq.~(\ref{3-4}) gives
the well defined infinite volume limit of the force, to 
first order in the displacements, generated by the system composed
both by the particles and the uniform negative background density.  Note
that Eq.~(\ref{3-4}) can be interpreted as either the force on the
particle at $\bu(0)$ due only to the other particles contained in the
sphere $S[r;\bu(0)]$ (with $r\rightarrow\infty$), 
with no contribution from the background due to symmetry reasons, 
or the sum of the force on the particle at $\bu(0)$ coming
from both the other particles contained in the sphere $S(r;0)$ (with
$r\rightarrow\infty$) and the
background in the same sphere. 

\section{Small displacements variance of the force}
\label{sec5}
We now turn to the problem of the variance of the force
$\left<F^2\right>$.  Considering Eq.~(\ref{3-4}), one might think 
that, if the variance of the displacements $\overline{u^2}\ll\ell^2$ 
then the right hand side of Eq.~(\ref{3-4}) can be used to
evaluate the variance of the force.  This is not, however,
as we will see, always true. In fact, we will show below, 
if $p(\bu)>0$ for $u>\ell/2$ in such a way to permit at 
least a pair (and therefore an infinite number of
pairs) of particles of the SL to be found arbitrarily
close to one another, then $\left<F^2\right>$ diverges due to the
divergence of the pair interaction for vanishing separation.  This
clarifies the meaning and the validity of this ``small displacement'' 
approximation given by Eq.~(\ref{3-4}): in order to use it to
calculate $\left<F^2\right>$, it is not sufficient to 
have $\overline{u^2}\ll \ell^2$. It is
instead necessary (and sufficient) that all the displacements are
smaller than half the elementary lattice cell size $\ell$ (i.e. that
the support of $p(\bu)$ be completely contained in the elementary lattice
cell) so that each particle has a positive minimal
separation from its first nearest neighbor.

In this section we will suppose that this is the case, i.e., displacements
are limited to a region contained in the elementary lattice cell.  Given
this hypothesis, because of the mutual statistical independence of the
displacements applied to different particles, the average quadratic
force acting on each particle, to order $\overline{u^2}$, is then
simply:
\begin{widetext}
\be 
\label{3-5}
\la F^2\ra \simeq \la |\bF^{(l)}|^2\ra=\la |\bF_A^{(l)}|^2\ra + 
\la |\bF_S^{(l)}|^2\ra
= \left(\frac{4\pi}{3}n_0\right)^2\overline{u^2}
+\lim_{r\rightarrow \infty} \sum_{\bR\ne 0}^{R\le r} \frac{\la
\left|\bu(\bR)-3[\bu(\bR)\cdot 
\hat\bR]\hat\bR\right|^2\ra }{R^6}\,.
\ee
\end{widetext}
If, moreover, we assume that not only the displacements of different
particles, but also the different components of the displacement of a
single particle are uncorrelated, it is simple to show that, 
for the class of lattices considered, we have
\be
\label{sym_c} 
\la\left|\bu(\bR)-3[\bu(\bR)\cdot\hat{\bR}]\hat{\bR}\right|^2\ra 
=2\la |\bu(\bR)|^2\ra=2\overline{u^2}\,.
\ee
With these hypotheses we can therefore write
\be 
\la |\bF^{(l)}|^2\ra= \left[\left(\frac{4\pi}{3}n_0\right)^2+
2\sum_{\bR\ne 0}\frac{1}{R^6}\right]\overline{u^2}\,,
\label{3-6}
\ee
where the sum is now over all the lattice vectors $\bR\ne 0$.
Performing the lattice sum in Eq.~(\ref{3-6}), for a simple 
cubic lattice, we find:
\bea
&&
2\overline{u^2}\sum_{\bR\ne 0}\frac{1}{R^6}=
12 \left( \frac{1}{\ell^3} \right)^2  \overline{u^2} 
\left(c_{1\mbox{nn}}+c_{2\mbox{nn}}+c_{3\mbox{nn}}+...\right) \nonumber
\\
&&\approx 
16.1 \left( \frac{1}{\ell^3} \right)^2 \overline{u^2}\,,
\label{F0}
\eea
where $c_{i\mbox{nn}}$ is the relative contribution to the sum
$\sum_{\bR\ne 0}\frac{1}{R^6}$ of the set of $i^{th}$ nearest
neighbor lattice sites of the origin $\bR=0$, normalized
such that $c_{1\mbox{nn}}=1$ (giving $c_{2\mbox{nn}}=1/4$ 
and $c_{3\mbox{nn}}=4/81$). With these approximations, we 
then find for a simple cubic lattice:
\be
\label{FSL}
\sqrt{\la F^2\ra}\simeq \sqrt{ \la |\bF^{(l)}|^2\ra}=\sqrt{
\langle |\bF^{(l)}_S|^2\rangle + \langle |\bF^{(l)}_A|^2
\rangle }=\alpha n_0U_0 \ee 
where $U_0=\sqrt{\overline{u^2}}$ and
\be \alpha \approx \sqrt{16.1 +
\left(\frac{4\pi}{3}\right)^2}  \approx 5.86\;.  
\ee 
Hence we can draw a first conclusion: while in the case of a
homogeneous Poisson particle distribution \cite{chandra} the
gravitational force acting on a given particle is dominated by the
first nearest neighbor, in the present case it is dominated by two
terms: the former is a global term $\bF_A$ due to the displacement
with respect to the rest of the system of the particle on which we 
calculate the force , and the latter, $\bF_S$, is mainly due to the set of
the first nearest neighbors lattice sites, which all lie at ``almost''
the same distance.  As shown explicitly below, because of the
symmetries of the initial lattice, the net gravitational force in a SL
is clearly very depressed, for small displacements, with respect to 
the single nearest neighbor contribution of the Poisson case with 
the same average density $n_0$.

\section{Useful results on the probability distribution of the force}
\label{sec6}

In the next section we will generalize to our case the method
introduced by Chandrasekhar \cite{chandra} for calculating the PDF
$P(\bF)$ of the gravitational force $\bF$ in a 3D homogeneous Poisson
particle distribution \cite{libro}.  As a starting point we briefly
report in this scetion Chandrasekhar's results for this latter case. 
The specific form of $P(\bF)$ in this case is called the 
{\em Holtzmark distribution}, and for this reason we will 
denote it $P_H(\bF)$. Subsequently, we give a brief presentation of the
exact derivation of the $P(\bF)$ in a $1D$ SL for a general power law
pair interaction.  Finally, we proceed to generalize Chandrasekhar's
method to the 3D SL by introducing some {\em ad hoc} approximations.
These results, together with those presented in previous sections on
the small displacement approximation, will provide us with a good qualitative
comprehension of the behavior of $P(\bF)$ in a 3D SL when the shape of
$p(\bu)$ is varied, and in particular when one passes from
displacements limited to an elementary lattice cell to larger maximal
displacements.

\subsection{Gravitational force PDF in a homogeneous 3D Poisson particle 
distribution}

\label{exact_pdf} 

In this subsection we give a brief account of the force PDF $P_H(\bF)$
for a three-dimensional homogeneous Poisson particle distribution
which, as aforementioned, is called the Holtzmark distribution.  The exact
mathematical derivation of this PDF can be found in \cite{chandra}.
One considers a homogeneous Poisson particle distribution
in a cubic volume with average number density $n_0$. Note
that, because of the total absence of correlations in the positions of
different particles, the PDF of the force/field at a point 
is the same whether the point is occupied or not by a particle.

While it is not possible to write an explicit expression for $P_H(\bF)$, 
is is possible for its FT
$A_H(\bq)={\cal F}[P_H(\bF)]$, i.e., for the characteristic function of
$\bF$ \cite{chandra}: 
\be 
A_H(\bq)\equiv e^{-n_0C_H(q)}=\exp \left[-n_0
\frac{4(2\pi q)^{3/2}}{15}\right]\,.
\label{char-poi}
\ee 
Note that this is a typical example of a characteristic function of a
{\em Levy stable distribution} \cite{stable} with exponent $3/2$. The
fact that $A_H(\bq)$ depends only on $q=|\bq|$ means that $P_H(\bF)$
depends only on $F=|\bF|$ (the particle distribution is statistically
isotropic). Since $A_H(\bq)$ is not analytic at $q=0$, $P_H(\bF)$ has
a power law tail at large $F$. Specificqlly, as $A_H(\bq)\simeq
[1-4n_0(2\pi q)^{3/2}/15]$, we have $P_H(\bF)\sim F^{-9/2}$. This shows
that moments $\la F^\alpha\ra$ of order $\alpha\ge 3/2$ diverge
 \footnote{It is possible to obtain these results noting that
 $-n_0C_H(q)$ is the {\em cumulant generating function} of the
 stochastic force $\bF$ which is non analytic at $q=0$ and with only
 one continuous derivative.}.  
In particular the variance of the force diverges. It is simple to see
that this is due to the small scale divergence of the pair
gravitational interaction together with the fact that particles can be
found arbitrarily close to one each other.  We then expect the same
large $F$ scaling behavior of $P(\bF)$ for a SL when the
support of $p(\bu)$ is larger than the elementary lattice size, as,
just as in the Poisson distribution, one can then find an infinite 
number of pairs of particles with members arbitrarily close to one 
each other. We show this below both
with the exact results in one dimension in Sect.~\ref{sec6b}, and with the
approximate approach {\em \`a la Chandrasekhar} in Sec~.\ref{sec7}.

We first recall the limiting behaviors of $P_H(\bF)$
which can be deduced directly from Eq.~(\ref{char-poi}):
\begin{subequations}
\bea
\label{h-ab} 
&&W_H(F) \sim \frac{4}{3\pi F_*^{3}} F^2 \;\;\;\mbox{ for} \;\;
F\rightarrow 0^+ \\ &&W_H(F) \sim \frac{15}{8} \sqrt{\frac{2}{\pi}}
F_*^{3/2} F^{-5/2} \;\;\mbox{ for} \;\; F\rightarrow \infty 
\label{h-ab-2}
\eea 
\end{subequations}
with
\[ F_*= 2\pi \left( \frac{4 n_0}{15} \right)^{2/3} \;,\] 
and where $W_H(F)=4\pi F^2P_H(\bF)$ is the PDF of the modulus of $\bF$.
The quantity $F_*$ can be seen as the typical force acting 
on a particle and is called the {\em normal
field} in \cite{chandra}.  It is also important to note that, in the
Poisson case for large values of $F$, the $W_H(F)$ is well
approximated by the PDF of the modulus of the force due only to the NN
particle \cite{libro}: 
\be
\label{poisson-force-nn}
W_{\text{NN}}(F) = 2 \pi n_0 F^{-5/2} \exp\left(-\frac{4 \pi n_0 F^{-3/2}}
{3} \right) \;.  \ee This means that the main contribution to the
force felt by a particle in a homogeneous Poisson distribution comes
from the first NN particle, and is due to the small distance
divergence of the pair gravitational interaction.

\subsection{Exact results for the 1D SL}
\label{sec6b}

Before introducing an approximate approach {\it \`a la Chandrasekhar} 
for the SL in 3D, we give some exact results obtained for the 
force PDF in an analogous 1D SL of particles
interacting via a power law interaction as
presented in \cite{andrea-1d}.

Let us consider a 1D SL, i.e., a set of $2N+1$ point-like particle
of unitary mass placed at the points $x_m=m\ell+u_m$ with $m=-N,...,N$
of the segment $[-L/2,L/2]$, where $\ell=L/(2N+1)$ is the lattice
spacing (and $n_0=1/\ell$ the average particle density), and $u_m$ is
the displacement of the $m^{th}$ particle from the lattice position
$m\ell$.  We assume that all the $u_m$ are extracted from the same PDF
$p(u)$ independently of one another, and that the particles
interact via the attractive pair force:
\[f(x)=\frac{x}{|x|^{\alpha+1}}\]
where $x$ is the particle separation and $\alpha>0$.

Therefore, the particle at $x_0=u_0$ feels the total force:
\be
F=\sum_{m\ne 0}^{-N,N}\frac{m\ell+u_m-u_0}{|m\ell+u_m-u_0|^{\alpha+1}}\,.
\label{1d-1}
\ee
Note that $F$ is not a sum of statistically independent
terms as each of these terms depend on two displacements, one of which
is always $u_0$.  By considering that the PDF of $x_m$ is simply given
by $p(x_m-m\ell)$, we can formally write the PDF $P(F)$ of the stochastic force
$F$ as \cite{andrea-1d}
\begin{widetext} 
\be
P(F)=\int..\int_{-\infty}^{+\infty}
\left[\prod_{m=-\infty}^{+\infty}dx_m\, p(x_m-m\ell)\right]
\delta\left(F- \sum_{m\ne
0}^{-\infty,+\infty} \frac{x_m-x_0}{|x_m-x_0|^{\alpha+1}}\right)\,,
\label{1d-2}
\ee 
\end{widetext}
in which we have taken directly the (symmetric) limit
$N\rightarrow\infty$ with $\ell$ fixed. 
By taking the FT in $F$ of Eq.~(\ref{1d-2}), one can write the 
characteristic function of the stochastic force as
\begin{widetext}
\be
\label{1d-6}
\tilde P(q)=\int_{-\infty}^{\infty}dx_0\,p(x_0)
\left[\prod_{n\ne 0}^{-\infty,+\infty}
\int_{-\infty}^{\infty}ds\,p(s+x_0-n\ell)
\exp\left(\frac{iqs}{|s|^{\alpha+1}}\right)\right]\,.
\ee
\end{widetext}
Through an exact analysis \cite{andrea-1d} of the small $q$ behavior of
Eq.~(\ref{1d-6}), 
one can distinguish essentially two classes of SL for what concerns the large 
$F$ behavior of $P(F)$:

(I) First class: No particle can be found arbitrarily close to any
other particle; i.e., the supports respectively of $p(u)$ and of
$p(u-n\ell)$, for all integer $n\ne 0$, have an empty overlap. 
Specifically this is the case if  $\exists \Delta\in
(0,\ell/2)$ such that $p(u)=0$ for $|u|>\Delta$, i.e., when the
support of $p(u)$ is all contained in one elementary lattice cell.  In
this case $P(F)$ vanishes at large $F$ for all values $\alpha>0$
faster than any negative power of $F$ and all the moments $\la F^n\ra$
are finite for any $n>0$. If moreover $p(u)=p(-u)$, $P(F)$ is not far
from a Gaussian with zero mean, even though there are deviations from
pure Gaussianity depending on the exact shape of $p(u)$. The finite
variance of $F$ in this case is given by: 
\begin{widetext}
\bea
&&\frac{\la F^2\ra}{2}=\sum_{l=1}^{+\infty}\sum_{n=1}^{+\infty}
\left[\left<\left<(u-x_0+n\ell)^{-\alpha}\right>_u
\left<(u-x_0+l\ell)^{-\alpha}\right>_u\right>_{x_0}
-\left<\left<(u-x_0+n\ell)^{-\alpha}\right>_u
\left<(u+x_0+l\ell)^{-\alpha}\right>_u\right>_{x_0}\right]\nonumber\\
&&+\sum_{n=1}^{+\infty}\left<\left<(u-x_0+n\ell)^{-2\alpha}
\right>_u-\left<(u-x_0+n\ell)^{-\alpha}\right>_u^2\right>_{x_0}
\label{1d-7}
\eea
\end{widetext}
where simply $\la(...)\ra_u=\int_{-\infty}^{+\infty} du\, p(u)(...)$
and $\la(...)\ra_{x_0}=\int_{-\infty}^{+\infty} dx_0\, p(x_0)(...)$.
Thus the force variance is composed of two distinct contributions: (i)
the double sum which is determined basically by the
fluctuations created by the stochastic displacement $x_0$ of the
particle initially at the origin on which we evaluate the force (in
this term the average over $u$, i.e., over the sources, plays the role of
a smoothing); (ii) the single sum which is instead
mainly due to the fluctuations in the displacements $u$ of all the
sources of the force (in this term only the average over $x_0$ is
a smoothing operation).  This coincides qualitatively with what
we have seen in Eq.~(\ref{3-6}) for the three dimensional case with an
approximate calculation.  The analogy with Eq.~(\ref{3-6}) can be made
stronger with the hypothesis of small displacements, i.e.,
$\overline{u^2}/\ell^2\ll 1$ where, in analogy with 3D, we have called
$\overline{u^2}=\int_{-\infty}^{+\infty} du\, p(u)u^2$.  Keeping only
terms up to the second order in the random displacements, it is simple
to show that Eq.~(\ref{1d-7}) can be rewritten as: \be \la
F^2\ra\simeq \frac{2\alpha^2\overline{u^2}}{\ell^{2(\alpha+1)}}\left[
2\zeta^2(\alpha+1)+\zeta(2\alpha+2)\right]\,,
\label{1d-8}
\ee where $\zeta(x)$ is the Riemann zeta function. In this expression
the first contribution comes from the fluctuations of the position of
the particle on which we calculate the force, and the latter comes from
the fluctuations of the positions of the sources.
In particular by writing
\[\zeta(2\alpha+2)=\sum_{n=1}^{\infty}\frac{1}{n^{2\alpha+2}}\,,\]
it is simple to show that the $n^{th}$ term in the sum gives the 
relative contribution of the $n^{th}$ NN lattice sites of the origin
to the force for $\overline{u^2}/\ell^2\ll 1$.

(II) Second class: At least one pair of particles (and therefore an
infinite number of pairs) can be found arbitrarily close to one each
other; i.e., the supports respectively of $p(u)$ and of $p(u-na)$, for
at least an integer $n\ne 0$, have a finite intersection.  The simplest case 
in this class is when $\exists \epsilon>0$ such that
$p(u)>0$ for all $|u|<\ell/2+\epsilon$.  In this case it is possible
to show \cite{andrea-1d} that the large $F$ tail of $P(F)$ is
proportional to $F^{-1-1/\alpha}$.  More precisely \be P(F)\simeq
B\,F^{-1-1/\alpha}\mbox{ for }F\rightarrow\infty
\label{eq23a}
\ee
with
\be
B= \frac{1}{\alpha}\int_{-\infty}^{+\infty}dx_0\, p(x_0)
\sum_{-2u^*/\ell<n<2u^*/\ell}^{n\ne 0}p(x_0-n\ell)\,,
\label{eq23}
\ee where $u^*$ is such that $p(u)>0$ for $u<u^*$ and vanishes for
$u>u^*$ (and $u^*=+\infty$ for $p(u)$ with unlimited support).  Note that
the large $F$ exponent of $P(F)$ is independent of the details of
$p(u)$. Moreover it coincides with the exponent characterizing the
large $F$ tail of the 1D Levy PDF found for a
1D Poisson particle distribution \footnote{It is simply
found by rephrasing the Chandrasekhar method to the 1D
case.}, i.e., the 1D analog of the Holtzmark distribution
seen in the previous section. In this case the amplitude $B$ of the tail in
the SL is $\ell\int_{-\infty}^{+\infty}dx_0\, p(x_0)
\sum_{-2u^*/\ell<n<2u^*/\ell}^{n\ne 0}p(x_0-n\ell)<1$ times smaller
than the amplitude of the Poisson case which is simply $1/(\alpha
\ell)$ \cite{andrea-1d}.  We note also that, if
$u^*\gg\ell$ and $p(u)$ is smooth (i.e. approximately constant) on
the length scale $\ell$, we can approximate Eq.~(\ref{eq23}) with \be
B=\frac{1-p(0)\,\ell}{\alpha \ell}\,.
\label{eq23b}
\ee
This last approximated expression is again independent of the
details of $p(u)$ for $u\ne 0$. 

Intermediate cases between (I) --- rapidly decreasing $P(F)$--- and (II)
---Holtzmark-like power law tail of $P(F)$--- are possible only if
displacements are permitted exactly up to $|u|=\ell/2$ but not beyond this 
value. In this case, depending on how $p(u)$ behaves in the neighborhood
of $u=\pm \ell/2$, one can have intermediate large $F$ tail behaviors
of $P(F)$, e.g., a power law decreasing faster than $F^{-1-1/\alpha}$.

The results just outlined coincide qualitatively with those
we will now find using an approximate generalization of 
Chandrasekhar's approach to the case of a three-dimensional 
SL. We will see that this method gives
accurate predictions on the large $F$ behavior of 
$P(\bF)$ for all $p(\bu)$, even though the accuracy for the 
amplitude of this tail is good only in the limit of 
sufficiently large displacements.

\section{Approximate Chandrasekhar approach to $P(\bF)$ in the 3D SL}
\label{sec7}

We extend here the formalism developed in \cite{chandra} in a similar
way to what has been done in \cite{gauss-poisson} for particle distributions
generated by a Gauss-Poisson process.  As shown in this latter paper, for
spatially correlated point processes in which each point has the same
mass, it is possible to introduce an approximated PDF for the
gravitational (or Coulomb), force felt by each 
particle (identical charge) of the system and due to all
the others. The approximation consists in using only the
information carried by one and two-point correlation functions of the
number density field.

Let us consider the general problem of a statistically homogeneous
particle distribution in a cubic volume of side $L$ (and hence of
volume $V=L^3$) and mean number density $n_0$ (with $L\gg n_0^{-1/3}$).
As usual the microscopic density is given by Eq.~(\ref{2-1}) with $m=1$.  
We study the PDF of the total gravitational force on the particle at $\bx_0$
due to the other $N$ points in the system in the limit $V\rightarrow
\infty$ with $N/V=n_0>0$ fixed (taking, as explained above,
the limit symmetrically with respect to the point $\bx_0$)  
\footnote{Because of the stochastic nature of
the point process, $N$ can deviate from $n_0V$ by a quantity growing
slower than $V$, which thus does not affect the results we present 
which are given in the infinite volume limit.}.  For simplicity let us take a
coordinate system such that $\bx_0$ coincides with the origin. The other
$N$ particles occupy the positions $\bx_1, \bx_2,...,\bx_N$
respectively.  The force acting on the probe particle at the origin is
\be \bF=\sum^{N}_{i=1}\frac{\bx_i}{x_i^3}\,.
\label{eq1}
\ee 
Let $p_c^{(N)}(\bx_1,\bx_2,...,\bx_N)$ be the conditional
joint PDF of the positions of the $N$ other particles with respect to
the probe at the origin.
The approximation we use consists essentially in assuming the validity
of the following factorization
\be
p_c^{(N)}(\bx_1,\bx_2,...,\bx_N)=\prod_{i=1}^{N}p_c(\bx_i)\,,
\label{eq2}
\ee where $p_c(\bx)$ is the conditional PDF of the position $\bx$ of a
given  particle with the condition that the origin of coordinates
is occupied by another particle of the system.  We thus
approximate the system seen by the particle at the origin with an
inhomogeneous Poisson particle distribution with space dependent average
density proportional to $p_c(\bx)$. This hypothesis works well in the
Gauss-Poisson case \cite{gauss-poisson}, and we expect it not to be bad for
any particle distribution which does not differ too much from a
Poisson one, i.e., with a two-point correlation function
$\xi(\bx)$ which is short range (i.e. integrable) and with a
small amplitude. This means that in our case of a SL we expect that 
the approximation
will give a quantitatively good estimate of $P(\bF)$ only when the typical
displacement of a particle starts to be of the order of the lattice cell size.
In fact it is only in this case that the lattice Bragg peaks contribution
to the PS of Eq.~(\ref{ps_sl}) is strongly reduced and, consequently,
the amplitude of $\xi(\bx)$ is small enough.  
However we will see that even for smaller displacements, when the
force variance is finite, this approximation gives good quantitative
predictions about the large $F$ tail of $P(\bF)$, even though the
value of the force variance is accurate only for the case in which the
largest permitted displacement starts to approach the cell boundary.  This
means that when instead the maximal permitted displacement is much
smaller than the lattice cell size we keep the qualitative results we
present in this section but use for the variance of $\bF$
Eqs.~(\ref{3-5}) and (\ref{3-6}).

Directly from the definition of the two-point correlation function, 
we have
\be
p_c(\bx)=A[1+\xi(\bx)]\,,
\label{eq2b}
\ee where $A\simeq 1/V$ is the normalization constant, and $\xi(\bx)$
for any stationary point process is defined as the off-diagonal part
of the covariance function $\tilde\xi(\bx)$ [see Eq.~(\ref{off})].  In
fact $\left<n(\bx)\right>_p=n_0[1+\xi(\bx)]$ gives the average
conditional density of particles seen by point occupied by a system
particle \cite{libro} (where $\left<..\right>_p$ indicates a
conditional average).

Making these hypotheses, it is possible to write \cite{libro}, 
in the infinite volume limit, the PDF of $\bF$ as 
\begin{widetext}
\be
\label{eq3}
P(\bF)=\int \frac{d^3q}{(2\pi)^3}\,e^{i\bq\cdot\bF}\exp
\left\{ -n_0\int d^3x\, [1+\xi(\bx)](1-e^{-i\bq\cdot \bx /
x^3})\right\}\,.
\ee
It will be useful to rewrite this in the  form
\be
\label{eq3b} 
P(\bF)=\int \frac{d^3q}{(2\pi)^3}\,e^{i\bq\cdot\bF} 
\exp\left[- n_0 C_H(q)-n_0\int d^3x\, \xi(\bx)(1-e^{-i\bq\cdot \bx /
x^3})\right]\,,\ee 
\end{widetext}
where $C_H(q)=4(2\pi q)^{3/2}/15$,
with the multiplicative factor $-n_0$, is the cumulant generating
function for $\bF$ in the homogeneous Poisson case already considered in
Sect.~\ref{exact_pdf} [see Eq.~(\ref{char-poi})].  The function 
\bea 
\label{eq3bb}
&& A(\bq)\equiv \int d^3F\,
e^{-i\bq\cdot\bF}P(\bF) \\  
&&=\exp
\left\{ -n_0\int d^3x\, [1+\xi(\bx)](1-e^{-i\bq\cdot \bx /
x^3})\right\}
\nonumber
\eea 
is the characteristic function of the stochastic force $\bF$.  
We recall that the function 
\be
{\cal G}(\bq)=\ln A(\bq)= -n_0\int d^3x\, [1+\xi(\bx)](1-e^{-i\bq\cdot 
\bx /x^3})
\label{eq3bbb}
\ee 
is the {\em cumulant generating function} of the stochastic field $\bF$
\cite{gardiner}.  The cumulants (i.e. the connected parts of the
moments) of $\bF$ can be directly calculated by taking the derivatives
of this function at $q=0$. Therefore the small $q$ behavior of ${\cal
G}(\bq)$ describes the large $F$ properties of $P(\bF)$.  Since 
in Eq.~(\ref{eq3bb}) the small $q$ region corresponds to the 
small $x$ region of $\xi(\bx)$ we can say roughly 
that the large $F$ behavior of $P(\bF)$ is
basically determined by the small separation properties of the
particle distribution [and therefore on the small $x$ behavior of
$\xi(\bx)$].  We have already considered this aspect both for the
Chandrasekhar method for the homogeneous Poisson particle distribution
and for the exact results in $1D$.  For the homogeneous Poisson case
the off-diagonal part of the reduced correlation function is
$\xi(\bx)\equiv 0$, and Eq.~(\ref{eq3bb}) becomes consistently
Eq.~(\ref{char-poi}) which implies Eq.~(\ref{h-ab}).

\subsection{The large $F$ behavior of $P(\bF)$}
\label{sec7-1}

In order to simplify the calculations which follow, we make the assumption
that  $\xi(\bx)=\xi(x)$ even though, rigorously speaking, this is not the
case for our SL because of the underlying lattice symmetry even when 
$p(\bu)$ depends only on $u=|\bu|$. 
For the SL this can be seen as an approximation consisting in
substituting $\xi(\bx)$ in Eq.~(\ref{eq3b}) with its angular average. 
Assuming $\xi(\bx)=\xi(x)$ implies that $A(\bq)$ and
$P(\bF)$ depend respectively only on $q$ and $F$, 
and Eq.~(\ref{eq3bbb}) can be rewritten as 
\begin{widetext}
\be
\label{eq6}
{\cal G}(\bq)=- 4\pi n_0 \int_0^\infty dx\,x^2 [1+\xi(x)]\left[1-
\frac{x^2}{q}\sin\left(\frac{q}{x^2}\right)\right]=
-n_0\left\{C_H(q)+4\pi \int_0^\infty dx\,x^2 \xi(x)\left[1-
\frac{x^2}{q}\sin\left(\frac{q}{x^2}\right)\right]\right\}\,,
\ee
\end{widetext}
Let us now analyze how the shape of $\xi(x)$ determines the large
$F$ tail of $P(\bF)$. 
To do this the fundamental step is to study the small $q$ behavior 
of ${\cal G}(\bq)$.
In this respect we distinguish below three cases for the
choice of $p(\bu)$ with a continuous and convex support: 
(Sect. \ref{sec7-1-1} ---{\em large displacements})
it permits, at least in some directions, displacements of particles 
beyond the border of their elementary lattice cells, 
allowing, in this way, different particles to be found
arbitrarily close to one each other; (Sect. \ref{sec7-1-2} ---{\em
marginal displacements}) it permits, at least in some directions,
displacements exactly up to the border of the elementary lattice cell
and in no direction beyond it, allowing different particles to be
found arbitrarily close to one each other, but only marginally;
(Sect. \ref{sec7-1-3} ---{\em small displacements}) its support is all
contained in a internal region of the elementary lattice cell 
so that there is a finite lower bound on the distance between
any two particles.

As has been noted in the discussion above about the accuracy of
 the approximation (\ref{eq2})
and (\ref{eq2b}), we expect to obtain better and better
approximations for $P(\bF)$ the larger is the typical
particle displacement.

In all three cases, it is important to note that ${\cal G}(0)=0$.
Moreover, as a SL is ``superhomogeneous'', 
$\int d^3x \xi(\bx)=-n_0=-\ell^{-3}$.
We now treat one by one the above three cases.

\subsubsection{Large displacements} 
\label{sec7-1-1}

In this case it simple to show that $\xi(0)>-1$ and finite. 
In fact the average conditional
density $n_0[1+\xi(\bx)]$ has to converge to a positive constant for
$x\rightarrow 0$, as the large displacements permit couples
of particles to be found arbitrarily close to one another.
This is sufficient to show (see Appendix~\ref{appI}) 
that, up to the leading term at small $q$, we have  
\[{\cal G}(\bq)\simeq -\frac{4}{15}n_0[1+\xi(0)](2\pi q)^{3/2}\,,\]
which is of the same order
in $q$ of $C_H(q)$. 
By recalling that $A(\bq)=\exp[{\cal G}(\bq)]$ 
and performing the Fourier transform (\ref{eq3b}), 
we can simply derive at large $F$ that 
\[P(\bF)\simeq [1+\xi(0)]P_H(\bF)\,,\]
which can be rewritten in terms of the PDF of $F=|\bF|$, $W(F)=4\pi
F^2P(\bF)$, as 
\be 
W(F)\simeq [1+\xi(0)]W_H(F)\simeq
[1+\xi(0)]\frac{15}{8} \sqrt{\frac{2}{\pi}} F_*^{3/2} F^{-5/2}\,,
\label{poilike}
\ee 
where we have used Eq.~(\ref{h-ab-2}).
This means that in this case $P(\bF)$ presents the same large force
scaling behavior as that of the Holtzmark distribution, but with an amplitude
greater by a factor $1+\xi(0)$.  
Given that $n_0[1+\xi(0)]$ is the average conditional density
of other particles at zero distance,
it is simple to show that for a SL one can write
\[1+\xi(0)=\frac{1}{n_0}\sum_{\bR\ne 0}\int d^3u\, p(\bu)p(\bu-\bR)\,,\]
from which one finds the 3D analog of Eq.~(\ref{eq23}).
Note that for most choices
of $p(\bu)$ one has also $\xi(0)<0$ (i.e. the system is negatively
correlated at small scales). Therefore the amplitude of the tail will
usually be reduced with respect to that in the Holtzmark case.  In any case,
as for the Holtzmark distribution, all the generalized moments
$\left<F^\eta\right>$ diverge for $\eta\ge 3/2$.

\subsubsection{Marginal displacements} 
\label{sec7-1-2}

In this case $\xi(x)\simeq [-1 + B x^\beta]$ with
$B,\beta>0$ at small $x$. In fact, as displacements are permitted up
to exactly the cell boundary, the probability of finding a particle at
distance $x$ from another fixed particle must vanish continuously for
$x\rightarrow 0$.  By studying the small scale behavior of ${\cal G}(\bq)$
(see Appendix~\ref{appI}), 
we can conclude that for $q\rightarrow 0$ we have:
\begin{itemize}
\item ${\cal G}(\bq)\sim -q^{(3+\beta)/2}$\\ 
if $0<\beta<1$,
which implies $W(F)\sim F^{-(5+\beta)/2}$ at large
$F$. In this case the variance $\left<F^2\right>$ is
thus divergent, but slower than in the Holtzmark case;
\item ${\cal G}(\bq)\sim q^{2}\log q$ \\ 
if $\beta=1$, implying $W(F)\sim F^{-3}$ (giving a
logarithmically divergent variance $\left<F^2\right>$);
\item ${\cal G}(\bq)\sim -q^{2}+o(q^2)\,,$\\ 
for $\beta>1$ where $o(q^2)$ is a
power vanishing faster than $2$ and including the main singular part of this
small $q$ expansion proportional to $q^{(3+\beta)/2}$ 
(with logarithmic corrections for $\beta$ integer larger than $1$).
This implies again $W(F)\sim F^{-(5+\beta)/2}$ at large $F$ (giving
a finite variance $\left<F^2\right>$).
\end{itemize}
Summarizing, we can say that in general $W(F)\sim F^{-(5+\beta)/2}$
with $\beta>0$ at large $F$ implying that all the moments
$\left<F^\eta\right>$ diverge for $\eta\ge (3+\beta)/2$.

\subsubsection{Small displacements} 
\label{sec7-1-3}

In this last case displacements are permitted up to a distance
$\Delta<\ell/2$. Consequently, there will be a positive minimal
distance $x^*=\ell -2\Delta$ between any two particles.
This implies that $\xi(x)=-1$ identically for $x<x^*$, as around any
particle there is a minimal empty region of radius $x^*$. 
Therefore Eq.~(\ref{eq6}) can be rewritten as 
\be 
{\cal G}(\bq)=-4\pi n_0\int_{x^*}^\infty dx\,x^2 [1+\xi(x)]\left[1-
\frac{x^2}{q}\sin\left(\frac{q}{x^2}\right)\right]\,.
\label{lnA-q}
\ee
The large $F$ behavior of Eq.~(\ref{eq3b}) is essentially 
determined by the small $q$
behavior of this function.  Since $x^*>0$, and $\xi(x)\rightarrow 0$
for $x\rightarrow\infty$, one can verify easily that the integral
in Eq.~(\ref{lnA-q}) can be expanded
in Taylor series to all orders in $q$ with finite
coefficients, i.e., ${\cal G}(\bq)$ and, consequently, $A(\bq)$ are 
analytic functions of $q$.  From Fourier transform 
theorems \cite{kolmogorov}, one can then infer that
$P(\bF)$ vanishes at large $F$ faster than any negative power of $F$,
i.e., $W(F)$ has all moments $\left<F^n\right>$ finite.
To the leading order in $q$, we can write 
\bea &&
\int_{x^*}^\infty dx\,x^2 [1+\xi(x)] \left[1-
\frac{x^2}{q}\sin\left(\frac{q}{x^2}\right)\right] \nonumber \\ &&
=\frac{q^2}{6}\int_{x^*}^{\infty} dx\,\frac{1+\xi(x)}{x^2}+O(q^4)\,.
\label{eq9}
\eea 
Therefore at small $q$ the characteristic function $A(\bq)$ has
the following behavior:
\be
A(\bq)=1-\frac{q^2\sigma_F^2}{6}+O(q^4)
\label{A-q}
\ee
where
\be
\sigma^2_F=4\pi n_0\int_{x^*}^{\infty}
dx\,\frac{1+\xi(x)}{x^2}
\label{s-f}
\ee 
is the approximation we obtain with this method for the 
variance $\la F^2\ra$ of the force $\bF$.  It is important to stress that
this formula for $\sigma^2_F$ is expected to apply to our SL only when
$\Delta/\ell$ approaches sufficiently the value $1/2$, i.e., when
displacements are large enough to make the amplitude of $\xi(x)$ small.
For smaller values of $\Delta$, Eqs.~(\ref{3-5}) and (\ref{3-6}) have
instead to be applied to calculate $\la F^2\ra$.

Note that the 3D isotropic
(i.e. monovariate) and uncorrelated Gaussian PDF reads
\be
P_G(\bF)=\left[\frac{3}{2\pi \sigma_F^2}\right]^{3/2}
\exp\left(-\frac{3F^2}{2\sigma_F^2}\right)
\label{P-gauss}
\ee
and has a characteristic function
\be
A_G(\bq)=\exp\left(-\frac{q^2\sigma_F^2}{6}\right)\,,
\label{P-gauss2}
\ee 
which has the same second order small $q$ expansion as that of
Eq.~(\ref{A-q}) [i.e. the same lowest order cumulant generating 
function ${\cal G}(\bq)$].  
Therefore we can say that in our case, if $\Delta$ is
not much smaller than $\ell/2$, $P(\bF)$ is approximately given by the
3D Gaussian (\ref{P-gauss}) where $\la F^2\ra$ is approximated by 
$\sigma_F^2$ which is given by the covariance function via Eq.~(\ref{s-f}).
Note, however, that Gaussianity is only approximate. In fact terms
of order equal or higher than $q^4$ from the expansion of
Eq.~(\ref{lnA-q}) are in general not vanishing differently from the case
of Eq.~(\ref{P-gauss}) where the cumulant generating function is a simple
quadratic function of $\bq$.  Instead, the fourth order term of the 
Taylor expansion of Eq.~(\ref{lnA-q}) can be written in general as
$4\pi n_0c_4q^4$ 
(the quantity $(\pi/6) n_0c_4$ is the fourth {\em cumulant} of $\bF$) with
\[
c_4=\frac{1}{5!}\int_{x^*}^{\infty}dx\frac{1+\xi(x)}{x^6}\,.
\]
In order to evaluate how large is the deviation from pure Gaussianity
due to the fourth (and higher order) cumulant term,
one has to compare $4\pi n_0 c_4$ with $\sigma^4_F/72$.  In fact the
fourth order term of the Taylor expansion of the pure
Gaussian $A_G(\bq)$ (Eq.~(\ref{P-gauss2})) is $(\sigma^4_F/72)q^4$, while
in our case the fourth order term of $A(\bq)$ is
$(\sigma^4_F/72+4\pi n_0 c_4)q^4$. Therefore the quantity 
\be
\lambda_{NG}=\frac{288\pi n_0 c_4}{\sigma^4_F}=\frac{3}{20\pi n_0}
\frac{\int_{x^*}^{\infty}dx[1+\xi(x)]/x^6} {\left[
\int_{x^*}^{\infty}dx[1+\xi(x)]/x^2\right]^2}
\label{dev}
\ee 
is a good measure of the degree of the {\em non-Gaussianity} of
$P(\bF)$. The quantity $\lambda_{NG}$ is called in probability theory
the {\em kurtosis excess} \cite{abramowitz}.  It measures the
importance of the large $F$ tail with respect to the Gaussian case
with the same variance. When $\lambda_{NG}\ll 1$ (i.e.
$2\Delta/\ell\ll 1$) we can say that the deviation of $P(\bF)$ from
the Gaussian $P_G(\bF)$ in Eq.~(\ref{P-gauss}) is small, while,
on the other hand, if instead $\lambda_{NG}\simeq 1$, the deviation 
starts to be appreciable and the
large $F$ tail of $P(\bF)$ starts to be considerably fatter than 
that of $P_H(\bF)$, and finally for $\lambda_{NG}\gg 1$ (i.e. for
$\Delta\simeq \ell/2$) the Gaussian approximation is inappropriate 
and $P(\bF)$ starts to develop the power law tail described 
above for the cases of large and marginal displacements.
This deviation from Gaussianity (see below) clearly increases with
$\Delta$ and in general diverges when $\Delta$ approaches $\ell/2$: in
fact for this value all the moments higher than a given value
diverge. Therefore for $2\Delta\rightarrow \ell^-$ we expect to see
large discrepancies of $P(\bF)$ with respect to Eq.~(\ref{P-gauss}).

In general for small $(x-x^*)>0$ the covariance $\xi(x)$ in the present
case behaves as $[1+\xi(x)]=B(x-x^*)^\beta$ with positive $B$ and
$\beta$ depending on $p(\bu)$. By changing the value of $\beta$ (i.e.
in our case of a SL, by changing the scaling behavior of $p(\bu)$ in the
neighborhood of $u=\Delta$), we can have or not a diverging behavior
of $\sigma_F^2$ and $c_4$ when $x^*\rightarrow 0^+$, that is when
$\Delta\rightarrow (\ell/2)^-$, and displacements are permitted up to
exactly the lattice cell boundaries. However in general we can say
that the deviation from pure Gaussianity given by $\lambda_{NG}$
increases with $\Delta$ and diverges at $\Delta=\ell/2$ for $\beta<5$.

For what concerns the approximate 
variance $\sigma_F^2$, it is simple to show that
if (i) $\beta<1$ this quantity diverges as $x^{*\,\beta-1}\sim
(\ell-2\Delta)^{\beta-1}$, if (ii) $\beta=1$ it diverges as
$-\log(\ell-2\Delta)$ (as we have already seen above in the case in
which $\Delta=\ell/2$ exactly), and if (iii) $\beta>1$ it converges to
a finite value. In a similar way it is simple to see that $c_4$
diverges as $(\ell-2\Delta)^{\beta-5}$ for $\beta<5$ (implying a large
$F$ tail slower than $1/F^7$, for $P(\bF)$), logarithmically for
$\beta=5$, and converges otherwise.

Finally Gaussianity (i.e. a PDF given by Eq.~(\ref{P-gauss})) is
almost exact when $\Delta\ll \ell/2$ \footnote{This can be seen more
rigorously noting that for this range of $\Delta$ the linear
expansion (\ref{3-5}) is valid. Therefore
each component of the force $\bF$ can be seen as the
sum of independent random variables with finite variance and
satisfying the {\em Lindeberg} condition \cite{lindeberg}. This
allows us to apply
the central limit theorem with good approximation to each
component of $\bF$ which, consequently, becomes a well defined
Gaussian variable in the infinite volume limit.}. However in this case the
approximation given by Eq.~(\ref{s-f}) for the variance $\la F^2\ra$
of $\bF$ in the SL is not a good one. In fact in this case
substituting such a SL with inhomogeneous Poisson particle
distribution with a radial density around the origin (where we
calculate the force) equal to the average conditional density of the
SL is a bad approximation as $\xi(x)$ acquires large values around the
Bragg peaks. Nevertheless, following also the results in $d=1$
presented above, in this case we can say again that $\bF$ is
approximately a 3D Gaussian variable [i.e. with a PDF given by
Eq.~(\ref{P-gauss})], but with $\la F^2\ra$ given by Eq.~(\ref{3-5}).

\subsection{Small $F$ behavior of $P(\bF)$}

In order to find the small $F$ behavior of $P(\bF)$, first of all
we note [see the first formula of Eq.~(\ref{h-ab})] that in the 
homogeneous Poisson case 
\[P_H(0)=4\pi\int_0^{\infty}dq\,q^2\exp [-n_0C_H(q)]\]
is finite, i.e., $W_H(F)\sim F^2$.
In our case, from Eqs.~(\ref{eq3b}) and (\ref{eq6}), 
$P(\bF)$ for $\bF=0$ is given by 
\begin{widetext}
\be
P(0)=\int d^3q A(\bq)=4\pi\int_0^{\infty}dq\,q^2
\exp \left\{-n_0C_H(q)
-4\pi n_0 \int_0^\infty dx\,x^2
\xi(x)\left[1-\frac{x^2}{q}\sin\left(\frac{q}{x^2}\right)\right]
\right\}\,.
\label{P-0}
\ee
\end{widetext} 
It is simple to verify that for any possible covariance functions
$\xi(x)$ the quantity $P(0)$ stays finite, i.e., again $W(F)\sim F^2$.
Roughly speaking, the more the particle distribution shows
anti-correlations, the larger will be the value $P(0)$, i.e., the
larger will be the probability of observing a small value of $F$. On
the contrary, the larger the positive correlations the smaller the
value of $P(0)$.  In particular, in our case of a randomly perturbed
lattice the system is {\em superhomogeneous}, that is $\int d^3x
\xi(\bx)=-1$, and therefore negative density-density correlations are
always more important than positive correlations.  This means that in
general, given the structure of Eq.~(\ref{P-0}) (and in particular
the fact that $\sin t/t<1$ for any $t>0$), $P(0)>P_H(0)$. Only in the
limit of random displacements in the whole system volume do we have
$P(0)\rightarrow P_H(0)$.  Moreover in general, the smaller are the
typical displacements, the larger will be the contribution of
anti-correlations to Eq.~(\ref{P-0}) and then the larger $P(0)$.

\section{Comparison with 
numerical simulations} 
\label{sec8}

In this section we compare our theoretical predictions for the
statistics of $\bF$ (specifically, the variance and the global PDF) 
we have given in the previous sections with numerical results for 
the same quantities obtained directly by computer simulations of 
the SL particle distribution with given $p(\bu)$.  
The paradigmatic example, on which we concentrate our numerical
analysis, is given by the case in which the three components $u_i$
with $i=1,2,3$ along the three orthogonal axes of the displacement
$\bu$ applied to the generic particle are statistically independent
and uniformly distributed in a symmetric interval, i.e., \be
p(\bu)=\prod_{i=1}^3 f(u_i)\,,
\label{p-u}
\ee
with
\be
\label{displa-SL}
f(u_i)=\left\{
\begin{array}{ll}
\frac{1}{2\Delta} & \mbox{for } 
|u_i| \le \Delta \\
&\\
0 & \mbox{otherwise}\,. 
\end{array}
\right.
\ee
In this case the average quadratic displacement is 
\[
\overline{u^2} = \frac{3}{2\Delta}\int_{-\Delta}^{+\Delta} x^2 f(x) dx 
= \Delta^2 
\]
or $\overline{u^2/\ell^2} =\delta^2$ in normalized units
$\delta = \Delta/\ell$.
The FT of $p(\bu)$, i.e., the characteristic function of the random 
displacements, is simply given by
\[\hat p(\bk)={\cal F}[p(\bu)]=\prod_{j=1}^3 \frac{\sin (k_j\Delta)}
{k_j\Delta}\,,\]
where $k_j$ is the $j^{th}$ component of $\bk$.
By using Eq.~(\ref{ps_sl}) it is then simple to verify that the PS 
${\cal P}(\bk)$ of the SL is given by:
\begin{widetext}
\be
{\cal P}(\bk)=\ell^3\left[1-\prod_{j=1}^3\frac{\sin^2 (k_j\Delta)}
{(k_j\Delta)^2}\right]
+(2\pi)^3\sum_{\bH\ne 0}\left[\prod_{j=1}^3\frac{\sin^2 (H_j\Delta)}
{(H_j\Delta)^2}\right]\delta(\bk-\bH)\,.
\label{ps_sl2}
\ee
\end{widetext}
This expression reduces at small $k$ to
\cite{GJS,displa} 
\be
\label{ps_sl_sk}
{\cal P}(\bk) = \frac{1}{3}\ell^3\Delta^2k^2 =
\frac{1}{3}\ell^5\delta^2k^2 \;, \ee which depends only on $k=|\bk|$,
i.e., mass (or number) fluctuations are statistically isotropic on
large scales even though the SL is not because of the underlying lattice
symmetry. In Fig.~\ref{sl-k2} the PS of the $1d$ analog of such a SL 
with $\delta<1/2$ is given. In this figure both the $k^2$ scaling behavior
at small $k$ and the modulation of the Bragg peaks of the initial lattice 
by the factor $|\hat p(k)|^2$ are clear.

For our particular choice of $p(\bu)$ it is possible to
calculate the inverse FT of Eq.~(\ref{ps_sl2}) to obtain, through
Eq.~(\ref{off}), the off-diagonal covariance function $\xi(\bx)$ for 
all values of $\Delta$. Let us call $n$ the integer part
of the ratio $4\Delta/\ell$. We can write
\begin{widetext}
\be
1+\xi(\bx)\!=\!\left(\frac{\ell}{2\Delta}\right)^3\prod_{j=1}^{3}
\left(\frac{|x_j|}{2\Delta}-1\right)\theta(2\Delta-|x_j|)
+\prod_{j=1}^{3}\!\left[1-\left(\frac{\Delta'}{\Delta}\right)^2
+\frac{\ell}{(2\Delta)^2}\!\sum_{m=-\infty}^{+\infty}
(2\Delta'-|x_j-m\ell|)\theta(2\Delta'-|x_j-m\ell|)\right]\,,
\label{xi-sh}
\ee
\end{widetext} 
where $\theta(x)$ is the usual Heaviside step function, and
$2\Delta'=(2\Delta-n\ell/2)$ if $n$ is even or zero and
$2\Delta'=[(n+1)\ell/2-2\Delta]$ if $n$ is odd. It is important to
notice that $0\le 2\Delta'<\ell/2$ for all $\Delta$ values. 
The function $\xi(\bx)$
given by Eq.~(\ref{xi-sh}) has in general a very complicated
oscillating form, with the exception, as shown below,
of the case in which $\Delta=m\ell/2$ exactly with
$m$ integer. However for a generic $\Delta$ we can say that it is 
continuous and is composed of two different contributions. The
former is given by the first product of Eq.~(\ref{xi-sh}) and comes
from the continuous (i.e. purely stochastic) part of the PS, and the
latter, given by the second product which is a lattice periodic function, 
comes from the modulated Bragg peaks part of the PS. 
For all the choices $\Delta=m\ell/2$
with integer $m$ the ``Bragg peaks contribution'' exactly
vanishes leaving only the first stochastic contribution: this means that
for these values of $\Delta$ the system becomes statistically 
translationally 
invariant. Note that for $\Delta<\ell/2$ we have
$\xi(\bx)=-1$ identically in the cube of side $|x_i|\le (\ell-2\Delta)$
for all $i=1,2,3$, meaning that the probability of finding a particle
in such a cube centered around any other particle is strictly zero
for these values of $\Delta$.

We can now draw the following general conclusions for all the possible
choices of $\delta=\Delta/\ell$. For simplicity we start with
the limiting case $\delta=1/2$, and then we analyze respectively the
cases $\delta<1/2$ and $\delta>1/2$.

\subsection{Large $F$ prediction for  $\Delta=\ell/2$}
\label{sec8-1}

For $\delta=1/2$, it is straightforward to find the exact
result 
\be
\xi(\bx)=-\prod_{j=1}^3(1-|x_j|/\ell) \theta(\ell-|x_j|)\,.
\label{unmezzo}
\ee 
Thus $\xi(\bx)$  is in this case non-vanishing, and negative, only 
in the cube $-\ell<x_j<\ell$ for all $j=1,2,3$.  
As mentioned above in this case
the lattice Bragg peaks of Eq.~(\ref{ps_sl2}) are completely erased
by the displacements and the system is statistically invariant
under translations. Expression (\ref{unmezzo}) at $x\ll \ell$ gives
\be 
\xi(\bx)\simeq -1+\sum_{j=1}^3\frac{|x_j|}{\ell}\,.
\label{small-x}
\ee 
Therefore the SL falls in the class of
Sect.~\ref{sec7-1-2} with $\beta=1$, i.e., with $P(\bF)\sim F^{-5}$ at
large $F$ [or equivalently$W(F)\sim F^{-3}$] and, from Eq.~(\ref{s-f}), 
logarithmically diverging variance $\sigma_F^2$. 
In order to have a more quantitative description of this case,
we mimic the anisotropic Eq.~(\ref{unmezzo}) with the following
isotropic $\xi(x)$:
\be
\xi'(x)=\left[-1+\left(\frac{\pi}{6}\right)^{1/3}\frac{x}{\ell}\right]
\theta\left[\left(\frac{6}{\pi}\right)^{1/3}\ell-x\right]\,,
\label{iso-app}
\ee 
i.e., an isotropic function with a linearly increasing behavior,
similar to the one in Eq.~(\ref{small-x}), from $\xi'=-1$ at $x=0$ to
$\xi'=0$ at the border and outside the sphere centered at $x=0$ with the same
volume $(2\ell)^3$ (i.e. radius $R=(6/\pi)^{1/3}\ell$) as that of 
the cubic region in which the function $\xi(\bx)$ in Eq.~(\ref{unmezzo})
increases from $-1$ to $0$.
By using this expression in Eq.~(\ref{eq6}), and for $q\ll \ell^2$, we obtain
\[A(\bq)\simeq \exp\left[2\left(\gamma q\right)^2
\ln\left(\gamma q\right)\right]=\epsilon(q)^{\epsilon(q)}
\,,\]
with $\gamma=(\pi/6)^{2/3}\ell^{-2}$ and $\epsilon(q)=\left(\gamma q\right)^2$.
By studying the inverse FT leading  from $A(\bq)$ to $P(\bF)$, 
and therefore to $W(F)$, we obtain at large $F$:
\be
W(F)\simeq 2\pi\left(\frac{\pi}{6}\right)^{1/3}\ell^{-4}F^{-3}\,.
\label{largeF}
\ee

\subsection{Large $F$ prediction for $\Delta<\ell/2$}
\label{sec8-2}

For $\delta< 1/2$ we can say that the SL falls in the class
of Sect.~\ref{sec7-1-3}, with $\beta=1$. It is simple to verify from
Eq.~(\ref{xi-sh}) that $\xi(\bx)=-1$ identically in the cube $|x_i|\le
(\ell-2\Delta)$ with $i=1,2,3$. As written in the previous section, in
this case $P(\bF)$ is expected to be rapidly decreasing at large $F$ and
with finite average quadratic force $\sigma_F^2=\left< F^2\right>$ as all the
higher order moments.  

More precisely, for $\delta\ll 1/2$, $P(\bF)$ is expected to be given,
to a good approximation, by Eq.~(\ref{P-gauss}) (i.e. Gaussian with 
a small kurtosis excess) with a variance $\sigma^2_F$ given by
Eq.~(\ref{3-6}) where $\overline{u^2}=\Delta^2$.  Instead for $\delta$
approaching $1/2$, i.e., $(1-2\delta)\ll 1$, we expect $\sigma_F^2$ to
be given, once a suitable isotropic approximation is introduced, by
Eq.~(\ref{s-f}).  A large kurtosis excess $\lambda_{NG}$ for this
range of $\delta$ is also expected, implying large deviations from pure
Gaussianity.  It is particularly interesting to study the diverging
behavior of $\sigma_F^2$ for $\delta\rightarrow(1/2)^-$ using
Eq.~(\ref{s-f}) to test the validity of this approximated analytical
result through comparison with measures from numerical simulations.
In order to apply Eq.~(\ref{s-f}) to our SL case we need an isotropic
approximation as good as possible for $\xi(\bx)$. First of all it is
important to note that (i) for $\bx$ just outside the cube $|x_i|\le
(\ell-2\Delta)$ with $i=1,2,3$, the function $\xi(\bx)$ grows
linearly, (ii) $\xi(\bx)$ grows up to the surface of the cube
$|x_i|\le 2\Delta\simeq \ell$ with $i=1,2,3$, (ii) outside the cube
$\xi(\bx)$ is function with the periodicity of the lattice with
amplitude at most of order $(1-2\delta)\ll 1$, and with zero mean on 
the period (i.e. on the elementary cell).  These observations, 
combined with the same argument leading to Eq.~(\ref{iso-app}), 
for $\delta=1/2$, permit one to approximate
$\xi(\bx)$ simply with 
\begin{widetext}
\be 
\xi'(x)\simeq \left\{
\begin{array}{ll}
-1&\mbox{ for }x<x^*\\
&\\
\left[\left(\frac{\pi}{6}\right)^{1/3}\frac{(x-x^*)}{2\Delta}-1\right]
\theta\left[\left(\frac{6}{\pi}\right)^{1/3}\ell-x\right]
&\mbox{ for }x\ge x^*
\end{array}
\right.\,,
\label{ao}
\ee 
\end{widetext}
where $x^*=(6/\pi)^{1/3}(\ell-2\Delta)$ is chosen so that 
$\xi'(x)=-1$ in the whole sphere around $x=0$ with the same
volume $8(\ell-2\Delta)^3$ as the cube where the exact 
$\xi(\bx)=-1$ identically,
and $(6/\pi)^{1/3}\ell$ is analogously the radius of the sphere with
volume $8\ell^3$ [i.e. the volume of the cube around $\bx=0$ outside
of which $\xi(\bx)$
is everywhere small and at most of order $(1-2\delta)$].  Clearly this is
a rough approximation to $\xi(\bx)$. However we will see that it
permits one to predict both the logarithmic divergence in $(1-2\delta)$ of
$\sigma_F^2$ and its order of magnitude.  In fact by using the
function (\ref{ao}) as $\xi(x)$ in Eq.~(\ref{s-f}), with $n_0=l^{-3}$,
we obtain the
following logarithmically diverging behavior for $\delta\rightarrow
(1/2)^-$ of $\sigma_F^2$: 
\be \sigma_F^2\simeq -4\pi\left(\frac{\pi}{6}\right)^{1/3}
\frac{\ln(1-2\delta)}{2\delta\ell^4}\,.
\label{ao2}
\ee 
Moreover, as discussed in the previous section, for values of
$\delta$ such that $(1-2\delta)\ll 1$ the kurtosis excess
$\lambda_{NG}$ is expected to be large, implying a large deviation
from Gaussianity of $P(\bF)$. This can be seen by using
Eq.~(\ref{ao}) in Eq.~(\ref{dev}), which gives for the
present case: 
\be 
\lambda_{NG}\simeq \frac{\delta}{400[\ln(1-2\delta)]^2}(1-2\delta)^{-4}\,.
\label{ao3}
\ee 
It is important to underline that this result is valid when Eq.~(\ref{ao})
is valid, i.e., when $\delta$ is not too far from the value $1/2$.
From Eq.~(\ref{ao3}) one can simply see that $\lambda_{NG}\approx 0.24$ for
$\delta=0.4$ and $\lambda_{NG}\approx 1.18$ for $\delta=0.44$. 
Moreover, beyond this value, $\lambda_{NG}$ 
diverges monotonously as $(1-2\delta)^{-4}$ for larger $\delta$. 
The right interpretation of this result is given directly by the
statistical meaning of the {\em kurtosis}:
by increasing $\delta$ in this range, $W(F)$ becomes more peaked and
with a large $F$ tail fatter and fatter than the 3D Gaussian 
distribution (\ref{P-gauss}) with the same variance $\sigma_F^2$.

\subsection{Large $F$ prediction for $\Delta>\ell/2$}
\label{sec8-3}

Since the approximated Chandrasekahr approach to the SL 
problem is more and more precise when $\delta$ increases,
we expect for $\delta>1/2$ a better quantitative agreement between the
analytic results and the numerical simulations than in the previous two
cases. In this range of displacements one has $\xi(0)>-1$ meaning
that on average a particle sees a density of other particles larger 
than zero at a vanishing separation, 
as each particle can be found arbitrarily close at least
to another particle.  
This implies that such a SL falls in the class of
Sect.~\ref{sec7-1-1} with a large $F$ tail of $W(F)$ with the same
scaling as that of the Holtzmark distribution but with a different amplitude
depending on the value of $\xi(0)$. This is clear from Eq.~(\ref{poilike})
which we rewrite here for convenience as 
\be
W(F)=[1+\xi(0)]W_H(F)
\label{poilike2}
\ee 
at large $F$ where $W_H(F)$ is given by Eq.~(\ref{h-ab-2}).  
As explained in Sect.~\ref{sec7-1}, as in the
homogeneous Poisson case, the statistically dominant contribution to
the force acting on a particle in this ccase comes from its nearest neighbor.
From Eq.~(\ref{xi-sh}) it is simple to find that: \be
1+\xi(0)=\left[1+\frac{\ell}{2\Delta}\left(\frac{\Delta'}{\Delta}
\right)-\left(\frac{\Delta'}{\Delta}\right)^2\right]^3
-\left(\frac{\ell}{2\Delta}\right)^3\,.
\label{xi-0}
\ee 
Therefore depending on the choice of $\Delta$ we can have both
$[1+\xi(0)]$ smaller  or larger than one, i.e., with a statistical weight
for large values of $F$ respectively smaller or larger than in the
homogeneous Poisson particle distribution, depending on whether the
probability of finding the nearest neighbor at very small distances is
smaller or larger than in the Poisson case.  However as the initial
lattice configuration presents negative density-density correlations
at small scales, for most of the choices of $\Delta$ one has $-1<\xi(0)<0$ and
therefore the large $F$ tail of $W(F)$ has a smaller amplitude than
$W_H(F)$. In general, by taking only the largest terms beyond one in
Eq.~(\ref{xi-0}), we can say that for large $\Delta$, the large $F$
ratio of $W(F)/W_H(F)$ approaches unity as $[1-O(\delta^{-2})]$ where
$O(x)\sim x$.

\subsection{Small $F$ predictions}
\label{sec8-4}

For what concerns the small $F$ behavior of $W(F)$ we have already
seen in the previous section that for all values of $\delta$ we have
$W(F)\simeq 4\pi P(0) F^2$ with the pre-factor $P(0)$ depending on
$\xi(x)$.  More precisely, for $\delta\ll 1/2$ we have just seen that
$P(\bF)$ coincides to a good approximation with the Gaussian
(\ref{P-gauss}). Therefore one has simply
$P(0)=\left[3/(2\pi\sigma_F^2)\right]^{3/2}$ where $\sigma^2_F$ is
given by Eq.~(\ref{3-6}). For higher values of $\delta$, when the
approximate Chandrasekhar approach starts to work, in order to find
$P(0)$, one should solve the integral (\ref{P-0}) where $\xi(x)$ is
some appropriate isotropic approximation of Eq.~(\ref{xi-sh}).
Clearly this is a task which it is very difficult or impossible
to perform analytically.  However 
for $\delta\ge 1/2$, i.e., when $W(F)$ is power
law at large $F$, one can adopt the following simple method to have a
rough approximation for the amplitude of the small $F$ tail of $W(F)$ and
therefore obtain a useful approximation of $W(F)$ for all values of
$F$ to be used to evaluate averages of arbitrary functions of $F$.
One assumes the following simple shape for $W(F)$: 
\be 
W(F)=\left\{
\ba {ll} 
AF^2&\mbox{for }F<F_0\\ BF^{-\alpha}&\mbox{for }F\ge F_0\,,
\ea \right.  
\label{simple-w}
\ee 
where respectively, as found above, $\alpha=3$ and
$B=2\pi(\pi/6)^{1/3}\ell^{-4}$ for $\delta=1/2$, and $\alpha=5/2$ and
$B=2\pi[1+\xi(0)]\ell^{-3}$ [where we have used Eqs.~(\ref{poilike2})
and (\ref{h-ab})] with $\xi(0)$ given by Eq.~(\ref{xi-0}) for
$\delta>1/2$. In order to find $A$ and $F_0$ we impose the following 
two conditions: (i) small $F$ and large $F$ tails take the same value 
at $F=F_0$, (ii) normalization of $W(F)$, i.e., $\int_0^\infty dF\,W(F)=1$.
The first condition implies 
\[AF_0^2=BF_0^{-\alpha}\,,\] 
while the second (normalization) condition gives the equation:
\[\frac{A}{3}F_0^3+\frac{B}{\alpha-1}F_0^{1-\alpha}=1\,.\]
These two equations can be solved to give:
\be
\left\{
\ba {l}
A=BF_0^{-2-\alpha}\\
F_0=\left[\frac{3(\alpha-1)}{B(\alpha+2)}\right]^{1/(1-\alpha)}\,.
\ea
\right.
\label{A-F_0}
\ee
In particular it is important to note that for $\delta>1/2$ 
these formulas imply that 
\be
\left\{
\ba{l}
F_0\sim [1+\xi(0)]^{2/3}\\
A\sim [1+\xi(0)]^{-2} \,,
\ea
\right.
\label{A-F_0-2}
\ee
where again $\xi(0)$ is given by
Eq.~(\ref{xi-0}).

\subsection{Comparison with numerical simulations}
\label{sec8-5}

To test the above analytical results, we have generated 
numerically several simple cubic SL, with fixed $\ell$ (for
simplicity we have chosen $\ell=1$, i.e., $n_0=1$) and 
different values of $\Delta$ in order to study 
$\sigma_F^2$ and $W(F)$ in a wide range going from 
$\delta\ll 1/2$ to $\delta >1/2$.  We expect then to
see the transition of $P(\bF)$ from nearly Gaussian for small $\delta$
to nearly Holtzmark for large values of $\delta$ when the particle
distribution approaches the Poisson one.  For each chosen value of
$\delta$ we have evaluated the PDF $W(F)$ in the following way: for each
realization of the SL the force is evaluated on the ``central'' particle
(i.e. on the particle farthest from the boundaries of the system);
then $W(F)$ is evaluated as a normalized histogram over $10^5$
realizations.  The force $\bF$ on the central particle is
computed by using the Ewald sum method for lattice sums for the cases
$\delta<0.3$ (i.e. when the SL keeps clear lattice features) in order
to make this evaluation faster and precise. For larger values
of $\delta$, on the other hand, $F$ is given by the simple sum of 
the contributions coming from all other particles included in the 
largest sphere centered on the central particle.

In Fig.~\ref{fig4} we present the numerical results for $\sigma_F$ vs.
$\delta$ for $\delta<1/2$ compared with the theoretical prediction for 
small displacements given by Eqs.~(\ref{3-6}) and (\ref{F0}). The agreement is 
excellent up to $\delta\simeq 0.2$.  Beyond this value $\sigma_F$ increases
faster than the theoretical prediction for small displacements
$\delta\ll 1/2$, and starts to show the diverging behavior for
$\delta\rightarrow (1/2)^{-}$ as predicted by Eqs.~(\ref{s-f}) and
(\ref{ao2}).

\bef
\includegraphics[height=6.5cm,width=8.5cm,angle=0]{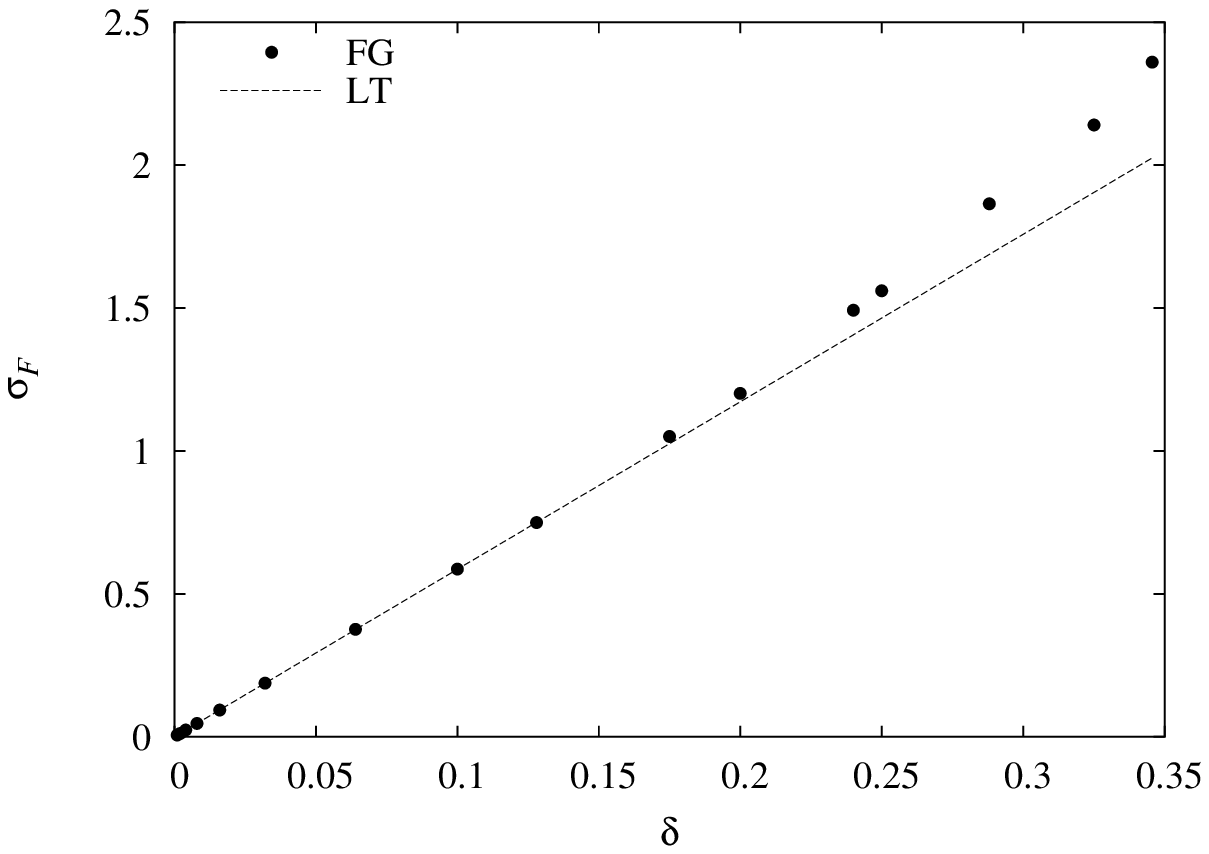}
\caption{Behavior of $\sigma_F$ vs. $\delta=\Delta/\ell$, for
$p(\bu)$ given by Eqs.~(\ref{p-u}) and (\ref{displa-SL}) with
$\delta<1/2$. For $\delta\le 0.2$ Eqs.~(\ref{3-6}) and (\ref{F0})
(dashed straight line), which are valid only for $\delta\ll 1/2$,
applies very well and the agreement with numerical results (circles)
is excellent up to approximately $\delta\simeq 0.2$. For larger values
the actual values of $\sigma_F$ starts to increase faster than this
simple linear prediction and the approximation {\em \`a la
Chandrasekhar} given in Sect.~\ref{sec7} starts to work as shown well
by the next figure.
\newline
\label{fig4}}
\eef 

This point is shown better by Fig.~\ref{fig5} where, in order to
show the logarithmically diverging behavior of $\sigma^2_F$ 
when $\delta\rightarrow (1/2)^-$, we
have plotted the numerical results for $\sigma^2_F$ vs.
$(1-2\delta)$, with a logarithmic scale for the latter, for
$0.4<\delta<0.499$ (and choosing as above $\ell=n_0=1$). Indeed, if the
approximated theoretical prediction Eq.~(\ref{ao2}) of a logarithmic
divergence is right, this should give a straight line.
This prediction is verified by the numerical simulations, 
albeit with a pre-factor in the logarithm which is smaller than 
the theoretical one. This discrepancy can be explained by the 
strong approximations
adopted in Sect.~\ref{sec7} to obtain Eq.~(\ref{ao2}). 
\bef
\includegraphics[height=6.5cm,width=9cm,angle=0]{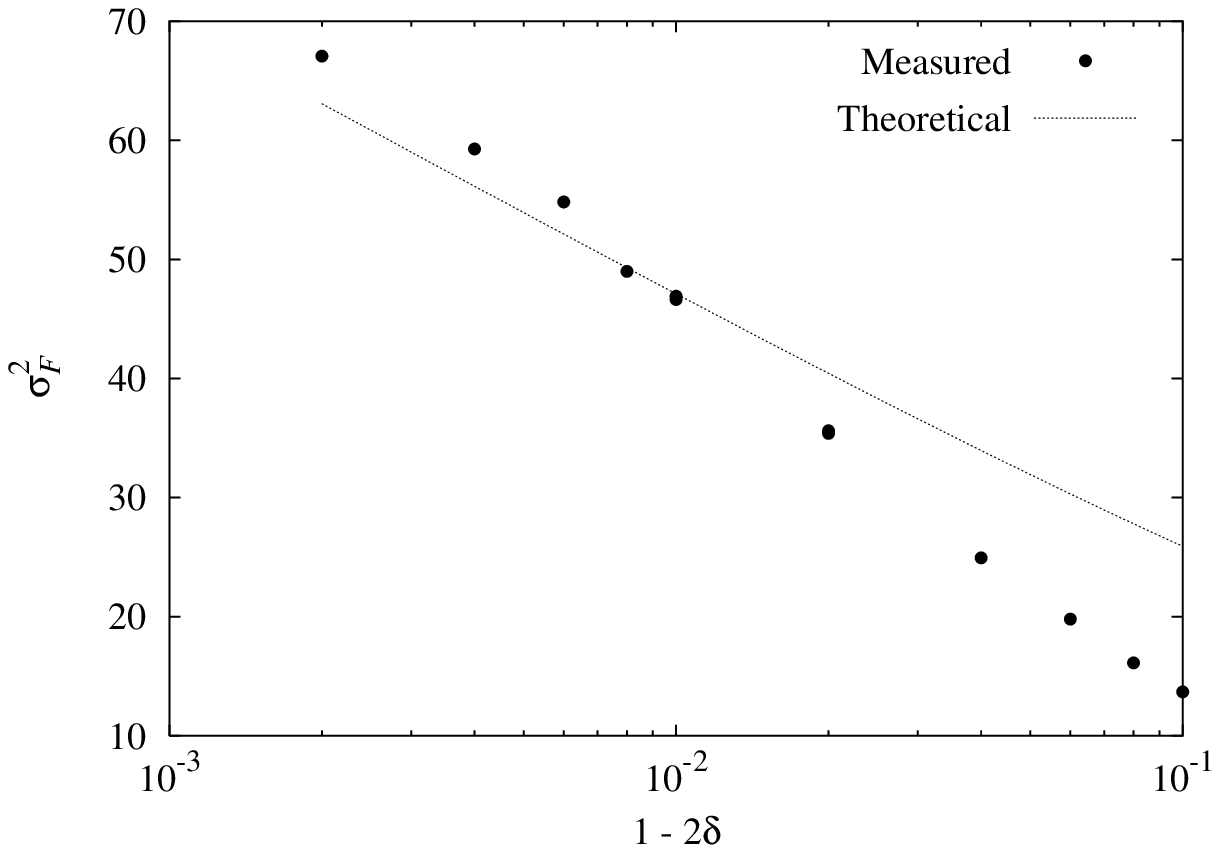}
\caption{Numerical results (circles, and best fit given by the dashed
line) vs. the approximated theoretical prediction (continuous
line) Eq.~(\ref{ao2}) obtained in Sect.~\ref{sec7}, with $\ell=n_0=1$.  
The numerical results, as predicted by Eq.~(\ref{ao2}), show a logarithmic
divergence of $\sigma_F^2$ in $(1-2\delta)$. However the slope is
about $20\%$ smaller than the approximated theoretical
prediction. Considering that this theoretical result is obtained by a
strong approximation (which becomes accurate only for larger values of
$\delta$), which consists in mimicking the SL with a inhomogeneous Poisson
particle distribution with radial density equal to the average
conditional density in the SL, we consider this to be a good
result.
\label{fig5}}
\eef 
****
In Fig.~\ref{fig6} we report the comparison for the PDF $W(F)$ between
the Gaussian theoretical prediction Eq.~(\ref{P-gauss}) and 
$\sigma_F^2$ as given by the linear approximation 
$\la|\bF^{(l)}|^2\ra$ of Eq.~(\ref{3-6}) for an
example of the case $\delta \ll 1/2$ (we have chosen $\delta=0.05$)
and the numerical results. 
\bef
\includegraphics[height=6.5cm,width=8.5cm,angle=0]{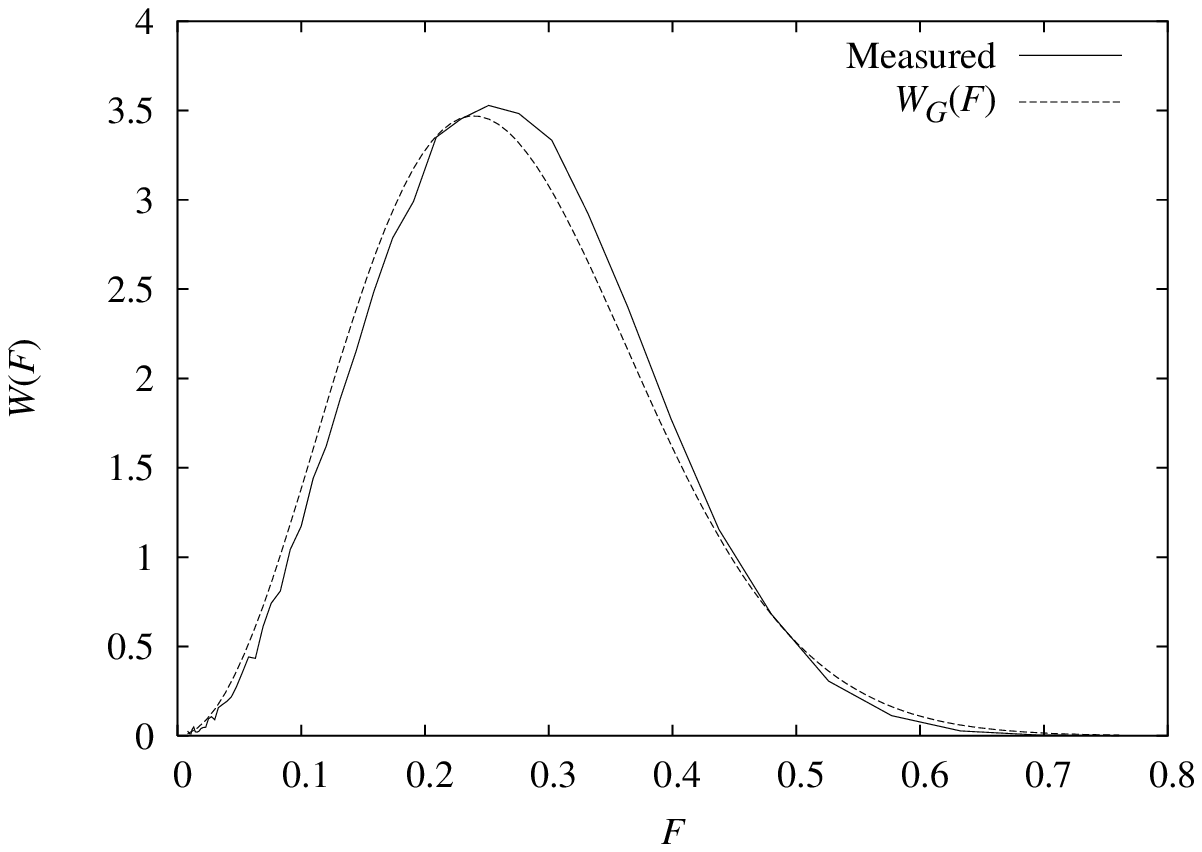}
\caption{$W(F)$ found numerically for a SL
with $\delta=0.05\ll 1/2$ and the theoretically predicted 
Gaussian distribution $W_G(F)=4\pi F^2 P_G(\bF)$
where $P_G(\bF)$ is defined by Eq.~(\ref{P-gauss}) with $\sigma_F^2$
given by $\la|\bF^{(l)}|^2\ra$ of Eq.~(\ref{3-6}).
The agreement between the two curves is very good.
\label{fig6}}
\eef 

In Fig.~\ref{fig6b}, on the other hand, we report the numerical 
evaluation of $W(F)$ for the cases $\delta=0.4$ and $\delta=0.45$ vs. 
the Gaussian $W_G(F)$ PDFs with the same variances. 
As theoretically predicted by Eq.~(\ref{ao3}), already for $\delta=0.4$
it starts to be evident that the actual $W(F)$ has a fatter large 
$F$ tail and a peak lower than that of $W_G(F)$. 
In fact for such values of $\delta$ the
kurtosis excess $\lambda_{NG}$ acquires a significantly
positive value. This discrepancy becomes even more clear
for $\delta=0.45$ for which $\lambda_{NG}\approx 2.1$.
In fact in this case the large $F$ tail starts to develop a power law
feature even though an exponential cut-off is still evident.

\bef
\includegraphics[height=6.5cm,width=8.5cm,angle=0]{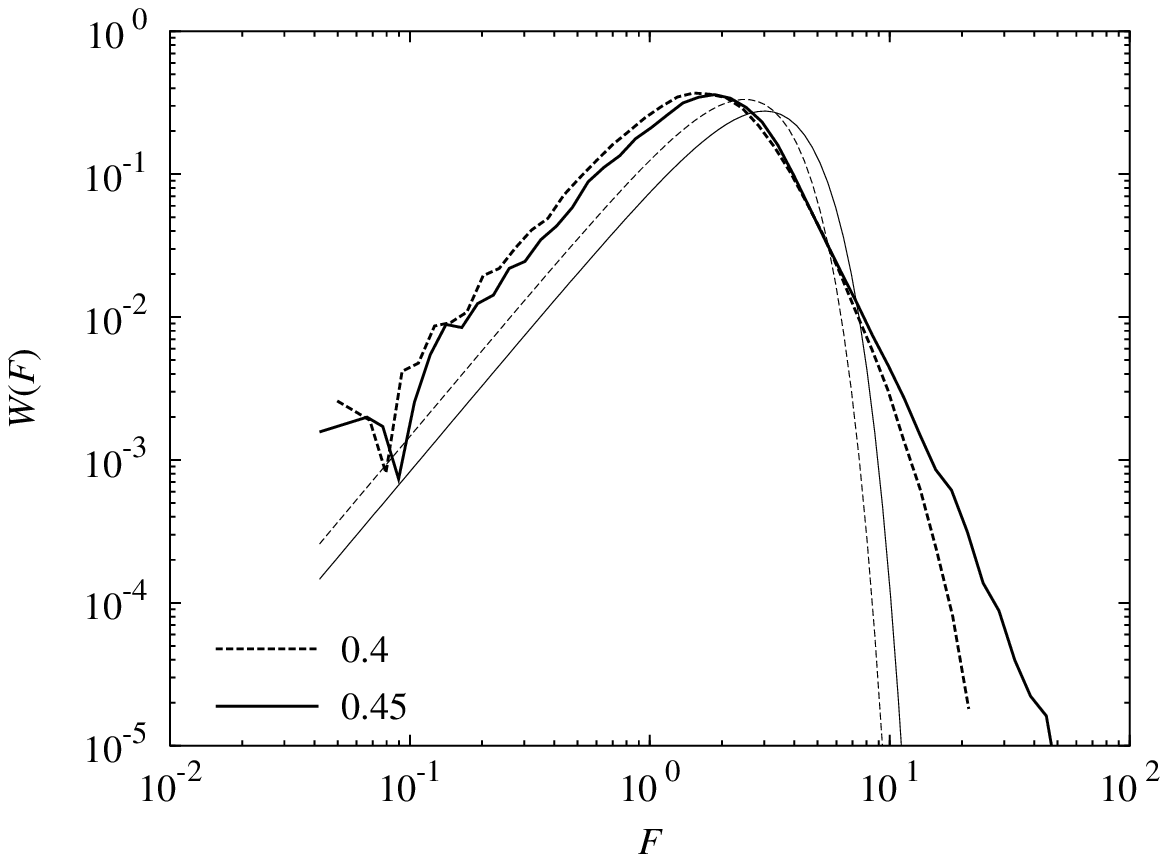}
\caption{Comparison between $W(F)$ found numerically for a SL
respectively with $\delta=0.4$ and $\delta=0.45$ and the  
Gaussian distributions $W_G(F)=4\pi F^2 P_G(\bF)$
with the same variances $\la F^2\ra$.
One observes that in both cases $W(F)$ has its peak at smaller $F$ and a more
significant large $F$ tail than the Gaussian approximation 
as predicted by Eq.~(\ref{ao3}) giving a
well defined positive kurtosis excess $\lambda_{NG}$.
The closer $\delta$ approaches to the ``critical'' value $1/2$,
the larger is this deviation.
\label{fig6b}}
\eef 

The ``critical'' case $\delta=1/2$ (i.e. where for the first time
$W(F)$ develops a power law large $F$ tail) is represented in
Fig.~\ref{fig7}, where the numerical $W(F)$ is compared with the
theoretical prediction for the large and small $F$ tails respectively
given by Eqs.~(\ref{largeF}) and (\ref{A-F_0}). Despite the roughness 
of the approximation, notably  for the small $F$ amplitude $A$, 
the agreement is very good.  
\bef
\includegraphics[height=6.5cm,width=8.5cm,angle=0]{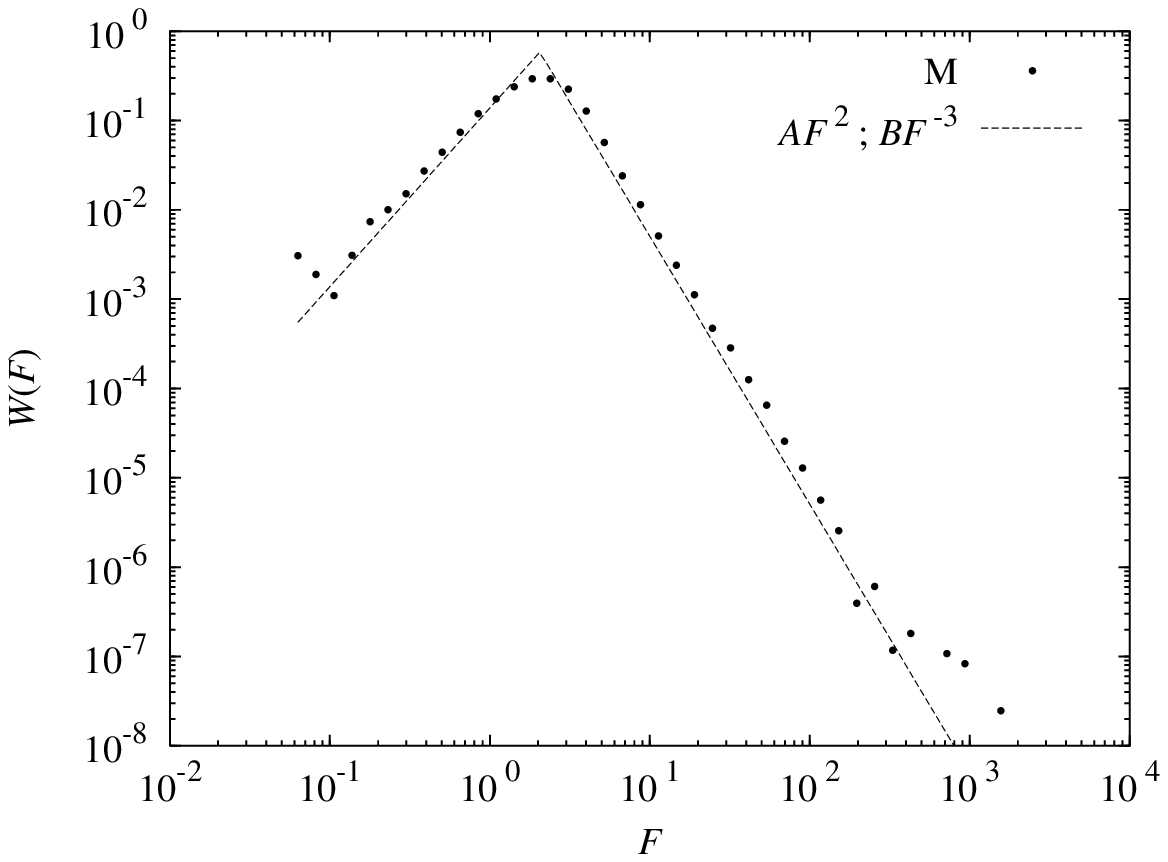}
\caption{Numerical $W(F)$ from simulations of computer realizations of
the SL with $\delta=1/2$. The large and small $F$ power law
approximations are given respectively by Eqs.~(\ref{largeF}) and
(\ref{A-F_0}).
\label{fig7}}
\eef

The Holtzmark-like case $\delta>1/2$ for $W(F)$ is represented in
Fig.~\ref{fig8} for the particular value $\delta=1$ of the shuffling
parameter. It is compared both with the exact Holtzmark distribution
obtained in a Poisson particle distribution with the same average
number density, and with the theoretical predictions for the large and small 
$F$ tails given respectively by Eq.~(\ref{poilike2}), with $\xi(0)$
given by Eq.~(\ref{xi-0}), and Eq.~(\ref{A-F_0}).
On the one hand we see that the $W(F)$ approximates quite well the 
exact Holtzmark one, confirming that the shape of $W(F)$ is mainly
determined by the small separation properties of the particle distribution.
On the other hand we see also that our theoretical approximation shows a
good agreement with simulations, although the small $F$ prediction 
is rougher than the large $F$ one. This is due to the very simple method
we have adopted in Sect.~\ref{sec8-4} to evaluate the amplitude of this tail
instead of calculating the more precise but difficult Eq.~(\ref{P-0}). 
\bef
\includegraphics[height=6.5cm,width=8.5cm,angle=0]{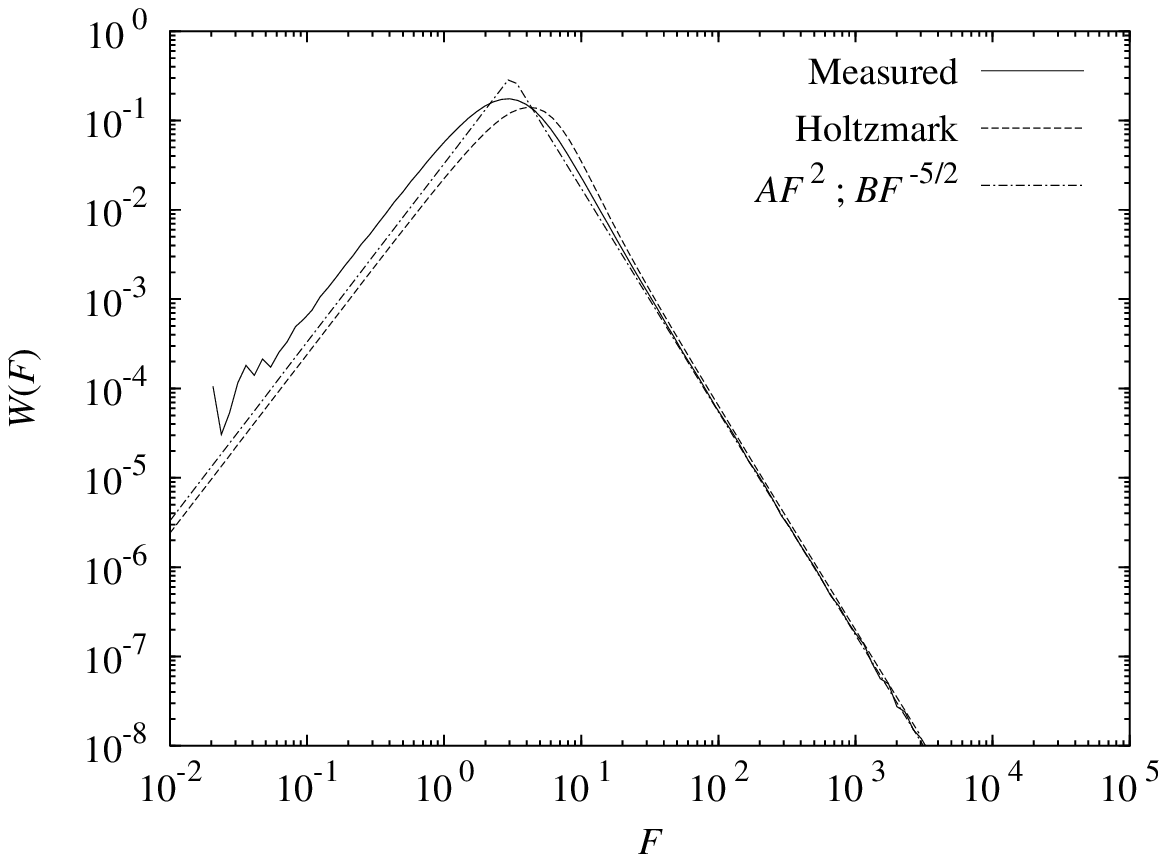}
\caption{Comparison between numerical $W(F)$ for the case $\delta=1$
with the exact Holtzmark distribution and the theoretical predictions
given by Eq.~(\ref{poilike2}), with $\xi(0)$
given by Eq.~(\ref{xi-0}), and Eq.~(\ref{A-F_0}).
\label{fig8}}
\eef 

Finally, as a further test of our theoretical
predictions, we have plotted the ratio $W(F)/W_H(F)$ giving a 
measure of the dependence on $\delta$ of the large $F$ tail of
$W(F)$ for a wide range of values $\delta>1/2$.
We have compared these values with the theoretical prediction given by 
Eq.~(\ref{poilike2}), with $\xi(0)$ given by Eq.~(\ref{xi-0}) as functions
of $\delta$. The agreement between numerical simulations and theory for
this quantity is impressive, particularly so given the non-monotonous
behavior of $\xi(0)$.
\bef
\includegraphics[height=6.5cm,width=8.5cm,angle=0]{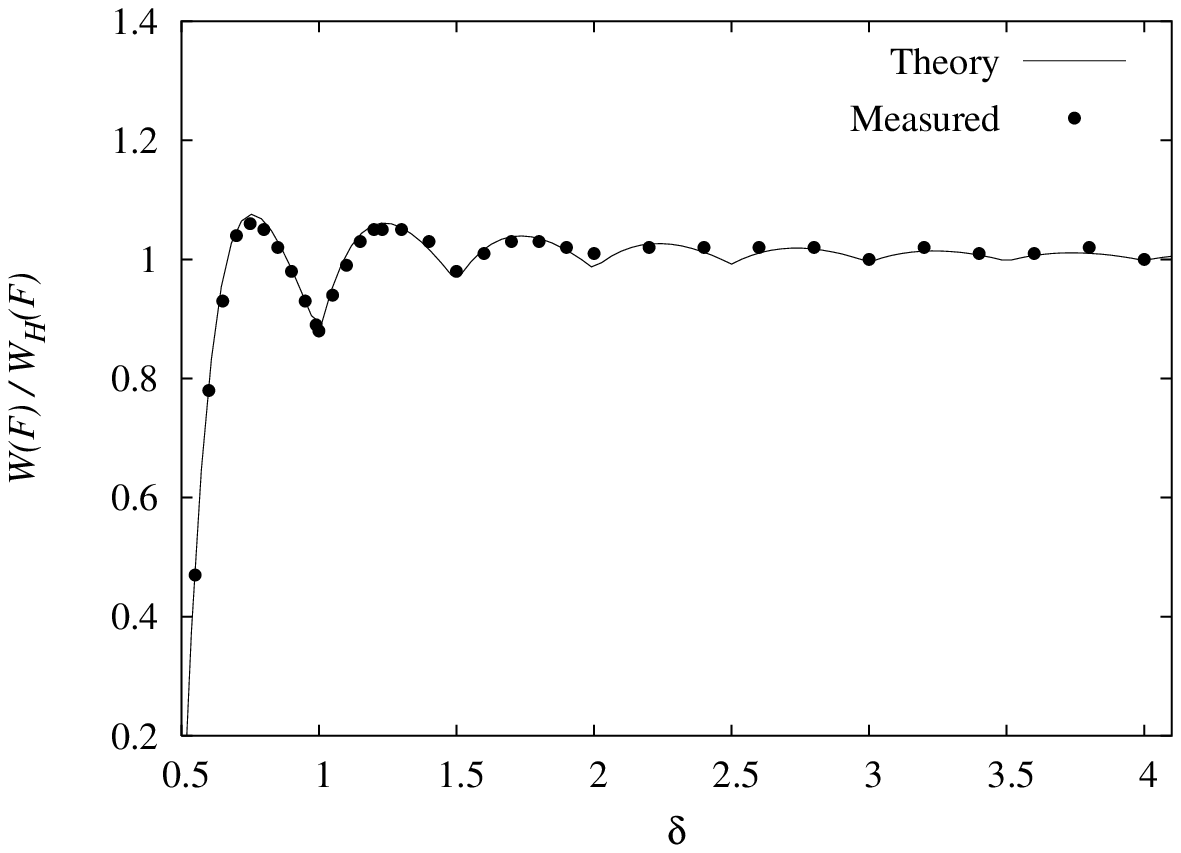}
\caption{ Plot of the $W(F)/W_H(F)$ for a wide range of
values $\delta>1/2$ compared with the theoretical prediction given by 
Eq.~(\ref{poilike2}), with $\xi(0)$ given by Eq.~(\ref{xi-0}) as functions
of $\delta$.
\label{fig9}}
\eef 

\section{Discussion and Conclusions} 
\label{sec9} 

In this paper we have presented a detailed study of the statistical
distribution of the total gravitational (or Coulomb) force acting on a 
particle belonging to a randomly perturbed lattice and due to the sum of the
pair gravitational interactions with all the other 
particles. 

In the first part of the paper we have studied the case in which the
displacements applied to the lattice particles to produce the
perturbed lattice are small. In particular we have analyzed the linear
expansion of the force in the displacements. Then we have seen that
if only displacements strictly smaller than half the lattice cell size are
permitted, this linear expansion can be used to calculate to a good
approximation the force variance. Otherwise this variance goes to
infinity, due to the small scale divergence of the pair interaction,
even though the average quadratic displacement is kept small.  We have
seen that in the case in which the force variance is finite, it can
be seen as the sum of two different terms: the former comes from the
small random displacements from the lattice position of the sources
keeping the particle on which the force is calculated fixed at its
initial lattice position, and the latter comes from the displacement
of this particle from the lattice position keeping at the same time
the sources at the initial lattice positions.  This second term can also
be seen as the contribution of the uniform negative background if it
is obtained by summing the contribution with respect to the original
lattice position of the particle feeling the force.

In the subsequent sections we have focused our attention on an
approximate extension to our case of the Chandrasekhar approach
leading to the Holtzmark PDF for the homogeneous
Poisson particle distribution.  In this way we have been able to find 
approximate expressions both for the force PDF and its characteristic
function (and the cumulant generating function) for all the range 
of typical displacements.  
We have seen that from a qualitative
point of view this functional prediction holds for the whole range of random
displacements, and that the agreement becomes quantitatively good for
typical displacements of the order of or larger than half the 
lattice cell size 
(i.e. when density-density correlations starts to be small and the
contribution of the Bragg peaks to the particle PS is strongly reduced).

All the above results have then been positively confirmed by a
direct comparison with numerical simulation of the system in
which the SL particle distributions are generated with a Monte Carlo
like method and the force probability distribution is numerically computed.

We have underlined that, in general, when $\delta$ starts to be of
the order of half the lattice cell size, i.e., when the minimal
permitted distance between particles vanishes, the force PDF is
dominated by the first NN contribution becoming very
similar to the Holtzmark distribution $W_H(F)$ even though at large
distances the SL particle distribution is still very different from a
homogeneous Poisson particle distribution. As has been noted, this is
due to the small scale divergence of the gravitational (or Coulomb)
pair interaction between particles. This suggests that, when
the minimal permitted distance between particles vanishes, the same
behavior for $W(F)$, both at small and large $F$, is expected to be 
found in all the spatial particle distributions sharing the same 
small scale correlation properties independently of the large scale 
features. 
This can be seen clearly in Eq.~(\ref{A-F_0-2}), where it is shown that
while the small and large $F$ exponent of $W(F)$ are universal
the amplitudes depend only on $\xi(0)$.

To conclude let us finally return briefly to comment on the 
applications of the results and methods we have just found.
They can be useful in various different contexts mentioned 
in the introduction, but we will discuss here only the primary 
application which has motivated our own study: the comprehension
of the dynamics of self-gravitating systems studied in cosmology.
In this context large numerical ``N-body'' simulations  of 
purely self-gravitating, essentially point-like 
\footnote{The smoothing generically introduced in the gravitational 
interaction is relevant only at distances much less than the 
initial inter-particle  separation.} particles are used to model
the evolution of a self-gravitating fluid. The probability 
distribution of the force on a given particle is a useful quantity 
to understand notably in considering (i) the early time dynamics 
and, more specifically, (ii) questions concerning 
the effects of discreteness in these simulations
(see \cite{bjsl02,prl}). In a forthcoming 
work \cite{bgjmslSL2006} we 
will report a full analysis of the dynamics of gravitational evolution 
of N-body simulations from precisely the SL initial conditions analysed
here, and the results given here will be directly applied in understanding 
these questions. Much can be understood from this study about the
case of real cosmological N-body simulations, in which the initial
conditions are lattices subject to small {\it correlated} perturbations.
In this respect we note that the methods developed here, notably the 
approximate generalisation of the Chandrasekhar method, can in principal
be generalised to such distributions. The present study of the SL is just
a first simpler starting point. Further the methods used here can be
seen as a first example for calculations of other statistical quantities
in such distributions of relevance in understanding
the dynamics of these self-gravitating systems at larger scales
and longer times, e.g., the probability distribution of the force
on the centre of mass of coarse-grained cells, the two-point correlation 
functions of gravitational force as a function of separation
etc..

\acknowledgments{
We thank the ``Centro Ricerche e Studi E. Fermi'' (Rome, Italy) for the use
of a super-computer for numerical calculations, the EC grant
No. 517588 ``Statistical Physics for Cosmic Structures" and the
MIUR-PRIN05 project on ``Dynamics and Thermodynamics of systems with
long range interactions" for financial supports.  M.\,J. thanks the
Instituto dei Sistemi Complessi for its kind hospitality during
October 2006. Finally we thank L. Pietronero for useful discussions 
and suggestions.
}
\bigskip

\begin{appendix}
\section{Small $q$ analysis of the cumulant generating function 
${\cal G}(\bq)$}
\label{appI}

In this appendix we provide some details of the small $q$ expansion
of the  cumulant generating function ${\cal G}(\bq)$ as defined in 
Eqs.~(\ref{eq3bbb}) and (\ref{eq6}) for $\Delta\ge \ell/2$.
As explained in Sect.~\ref{sec7}, for such values of $\Delta$ and sufficiently
small $x$ the correlation function $\xi(x)$ can be written as
\be
\xi(x)=\xi(0)+Bx^\beta+o(x^\beta)\,,
\label{ap1}
\ee
where $\xi(0)=-1$ for $\Delta=\ell/2$ and $-1<\xi(0)<+\infty$ for
$\Delta>\ell/2$, and $B, \beta>0$.
Let us suppose that the expansion (\ref{ap1}) is valid for 
$x<x_0$ with $x_0>0$, and rewrite the integral in Eq.~(\ref{eq6})
as the following sum of two integrals:
\begin{widetext}
\be
{\cal G}(\bq)=-4\pi n_0 \left\{\int_0^{x_0} dx\,x^2[1+\xi(x)]\left[1-
\frac{x^2}{q}\sin \left(\frac{q}{x^2}\right)\right]
+\int_{x_0}^{+\infty} dx\,x^2[1+\xi(x)]\left[1-
\frac{x^2}{q}\sin \left(\frac{q}{x^2}\right)\right]\right\}\,.
\label{ap2}
\ee
\end{widetext}
Now in the first integral we can use Eq.~(\ref{ap1}), while in the second 
one, assuming $q\ll x_0^2$ we can expand $\sin(q/x^2)$ in Taylor series.
Since $x_0>0$ and independent of $q$, and $\xi(x)$ vanishes for
$x\rightarrow+\infty$, it is now simple to show that the
second integral is of order $q^2$ at small $q$.
Let us call $I(q)$ the first integral, i.e.,
\[I(q)= \int_0^{x_0} dx\,x^2[1+\xi(x)]\left[1-
\frac{x^2}{q}\sin \left(\frac{q}{x^2}\right)\right]\,.\]
By using Eq.~(\ref{ap1}) we have:
\begin{widetext}
\be
\label{ap3}
I(q)=[1+\xi(0)]q^{3/2}\int_0^{x_0/\sqrt{q}}dt\,t^2
\left[1-t^2\sin(t^{-2})\right]
+Bq^{(3+\beta)/2}\int_0^{x_0/\sqrt{q}}dt\,t^{2+\beta}
\left[1-t^2\sin(t^{-2})\right]
+o\left[q^{(3+\beta)/2}\right]\,,
\ee
\end{widetext}
where we have changed variable to $t=x/\sqrt{q}$ in both
integrals.
In Eq.~(\ref{ap3}), the first integral converges in the limit
$q\rightarrow 0$ while the second one converges only for $0<\beta<1$, diverges
logarithmically for $\beta=1$, and diverges as 
$q^{(1-\beta)/2}$ for $\beta>1$.
Therefore we can conclude that for $\xi(0)>-1$, i.e., $\Delta>\ell/2$,
in the limit $q\rightarrow 0$ up to the dominant term we have:
\begin{widetext}
\be
\label{ap4}
{\cal G}(\bq)\simeq -4\pi n_0 I(q)\simeq -\left\{4\pi n_0 
[1+\xi(0)]\int_0^\infty dt\,t^2
\left[1-t^2\sin(t^{-2})\right]\right\}q^{3/2}=-n_0[1+\xi(0)]C_H(q)\,.
\ee
\end{widetext}
Instead for $\xi(0)=-1$, i.e., $\Delta=\ell/2$, the coefficient of the 
first integral of $I(q)$ vanishes, and the second one 
[considering also the second integral of Eq.~(\ref{ap2}) 
for $\beta>1$, which is also of order $q^2$] gives:
\be
\label{ap5}
{\cal G}(\bq)\propto \left\{
\ba {ll}
-q^{(3+\beta)/2}&\mbox{for }0<\beta<1\\ 
q^2\ln q&\mbox{for } \beta=1\\ 
-q^2&\mbox{for } \beta>1\,.\\ 
\ea
\right.
\ee
Note that even for $\beta>1$, differently from the case $\Delta<\ell/2$
the small $q$ expansion of ${\cal G}(\bq)$ contains a singular part
even though of order higher than $q^2$.

\end{appendix}

\end{document}